\documentclass[aps,prc,twocolumn,showpacs,floatfix,preprintnumbers,superscriptaddress,amsmath,amssymb,longbibliography]{revtex4-1}

\usepackage[colorlinks,citecolor=blue]{hyperref}
\usepackage[colorinlistoftodos]{todonotes}
\usepackage[normalem]{ulem}
\usepackage{color}
\usepackage[utf8]{inputenc}
\usepackage{bm}

\makeatletter
\def\@bibdataout@aps{%
\immediate\write\@bibdataout{%
@CONTROL{%
apsrev41Control%
\longbibliography@sw{%
    ,author="08",editor="1",pages="1",title="0",year="1"%
    }{%
    ,author="08",editor="1",pages="1",title="",year="1"%
    }%
  }%
}%
\if@filesw \immediate \write \@auxout {\string \citation {apsrev41Control}}\fi 
}
\makeatother

\newcommand{\rpn}{\textit{r} process}
\newcommand{\rpa}{\textit{r}-process}

\begin{document}

\title{Efficient method for estimation of fission fragment yields of {\rpa} nuclei}

\author{Jhilam Sadhukhan}
\affiliation{Physics Group, Variable Energy Cyclotron Centre, 1/AF Bidhan Nagar,
Kolkata 700064, India}
\affiliation{Homi Bhabha National Institute, Anushakti Nagar, Mumbai 400094,
India}

\author{Samuel A. Giuliani}
\affiliation{FRIB Laboratory,
Michigan State University, East Lansing, Michigan 48824, USA}

\author{Zachary Matheson}
\affiliation{FRIB Laboratory,
Michigan State University, East Lansing, Michigan 48824, USA}

\author{Witold Nazarewicz}
\affiliation{Department of Physics and Astronomy and FRIB Laboratory,
Michigan State University, East Lansing, Michigan 48824, USA}


\begin{abstract}
\begin{description}
\item[Background]
	More than half of all the elements heavier than iron are made by the
	rapid neutron capture process (or  {\rpn}). For very neutron-rich
	astrophysical conditions, such at those found in the tidal ejecta of
	neutron stars, nuclear fission determines the {\rpa} endpoint, and the
	fission fragment yields shape the final abundances of $110\le A \le
	170$ nuclei. The knowledge of fission fragment yields of hundreds of
	nuclei inhabiting very neutron-rich regions of the nuclear landscape is
	thus crucial for the modeling of heavy-element nucleosynthesis. 
\item[Purpose]	
	In this
	study, we propose a  model for the fast calculation of fission
	fragment yields based on the concept of shell-stabilized prefragments
	defined with help of the nucleonic localization functions.
\item[Methods]	
	To generate realistic potential energy surfaces and nucleonic localizations, we apply Skyrme  Density Functional Theory. The
	distribution of the neck nucleons among the two prefragments is obtained
	by means of a statistical model. 
\item[Results]	
	 We benchmark the method by
	studying the fission yields of $^{178}$Pt, $^{240}$Pu, $^{254}$Cf, and
	$^{254,256,258}$Fm and show that it satisfactorily explains the
	experimental data. We then make predictions for $^{254}$Pu and
	$^{290}$Fm as two representative cases of fissioning nuclei that are
	expected to significantly contribute during the {\rpa} nucleosynthesis
	occurring in neutron star mergers.
\item[Conclusions]	
The proposed framework provides an
	efficient alternative to microscopic approaches based on the evolution
	of the system in a space of collective coordinates all the way to scission. It can be used  to carry out global
calculations of fission fragment distributions across the {\rpa} region. 
\end{description}
 \end{abstract}


\maketitle

\section{INTRODUCTION} 
The predictive power of \rpa\ network calculations could
be improved by providing sound predictions of fission fragment yields, including
yields from spontaneous fission (SF), beta-delayed fission, and neutron-induced
fission. In particular, it has been long established that fission fragment
distributions play a crucial role in shaping the final \rpa\
abundances~\cite{horowitz2018,goriely2013, eichler2015, goriely2015a, vassh2019,
vassh2019a}, and recent studies suggest that uncertainties in fission yields
should be reduced in order to properly constrain the contribution of binary
neutron star mergers to the chemical evolution of \rpa\
elements~\cite{cote2017}. When it comes to realistic predictions, the Langevin
approach~\cite{sadhukhan2016,sierk2017,usang2019} has proven to be extremely
successful in terms of quantitative reproduction of fragment yields.
Unfortunately, full-fledged Langevin calculations in a multidimensional
collective space, based on the microscopic nuclear density functional theory (DFT)
input, are computationally expensive when it comes to large-scale theoretical
fission surveys. The same is true for the time-dependent 
calculations~\cite{simenel2018,scamps2015a,bulgac2018,bulgac2019,regnier2019,zhao2019} based on self-consistent approaches. Given the
computational cost of microscopic methods and the large number of fissioning
nuclei that are expected to contribute during the \rpa\ nucleosynthesis, network
calculations have mostly relied on simple parametrizations or highly
phenomenological models to determine fission yields.

To date, the models that aim at  global surveys of fission yields can be grouped
into three main categories \cite{schmidt2018}. The first group comprises the so-called Brownian
motion models, wherein the fission process is described as an overdamped motion
across potential energy surfaces obtained from microscopic-macroscopic
calculations~\cite{randrup2011,moller2015c,mumpower2019}. These models take into
account the dissipative dynamics required to describe the full distribution of
fission yields. To deduce  charge yields, additional assumptions are made such
as  a linear scaling of  mass distributions. 

Methods belonging to the second category, the scission-point
models~\cite{fong1953,Erba1966,wilkins1976}, are free from this limitation
because they treat charge asymmetry as a collective degree of freedom. However,
the scission-point approach relies on a static description of the fragments
along the scission configuration, thus neglecting fission dynamics. For recent
realizations of this model, see, e.g.,
Refs.~\cite{lemaitre2015,lemaitre2019,Andreev2006,Pasca2019}. Within this category,
another simplified static description is the random neck-rupture model~\cite{brosa1983,Ober1998} where the
location of the neck is decided randomly following an appropriate statistical weight. 

Finally, the third category are semi-empirical models, such as
GEF~\cite{schmidt2016} and ABLA~\cite{kelic2009}, which have been widely used in
\rpa\ calculations~\cite{eichler2015, mendoza2015, goriely2015, goriely2015a,
vassh2019}. These models turned out to be  extremely successful in describing
data close to stability, since their parameters have been fine-tuned to
reproduce observed fission fragments distributions. However, their applicability
to extrapolate far from stability is questionable and the related uncertainties
are difficult to asses.

To overcome these shortcomings, the present paper presents an efficient and
predictive model for fission fragment yields that is based on the microscopic
DFT input. There are two main assumptions behind our model: (i) the formation of
fission fragments is governed mainly by shell effects of the prefragments that
develop in the pre-scission region; and (ii) the final fragments are produced in
the scission region by a rapid distribution of neck nucleons, a process that is
statistical in nature. Both assumptions are based on results of previous
microscopic calculations~\cite{sadhukhan2016,sadhukhan2017,matheson2019}.
Indeed, by studying the nucleonic localization function (NLF) for deformed
configurations of fissioning nuclei, we observe that two distinct prefragments
form and then separate well before scission is
reached. This early development of the prefragments is a manifestation of the
freeze-out of single particle energies along the fission
path~\cite{negele1989,nazarewicz1993,sadhukhan2013,scamps2018a} as the system
tries to maintain its microscopic configuration to avoid level crossings. The
suggestion that the prefragment particle numbers play essential roles in
determining the fission fragment distribution can be traced back to strong shell
effects in prefragments in the semiclassical periodic-orbit
theory~\cite{Strutinsky1976,Arita2020}.

Within the proposed model, it is not necessary to calculate the full potential
energy surface (PES) up to the scission configuration, since  a simplified
estimation of the PES is sufficient to obtain the most-probable prefragment
configuration(s). Then, final fragment yields are generated by distributing the
neck nucleons into prefragments according to a statistical prescription wherein
the frequency of a particular fragmentation channel is decided with an
appropriate microcanonical probability~\cite{fong1953,Bondorf1995}. 

The paper is organized as follows. Our model is described in Sec.~\ref{Smodel}.
Section~\ref{Sresults}  contains the benchmark calculations of fission yields
and  presents results for  two representative cases of \rpa\ nuclei.
Section~\ref{Sconclusions} contains a summary and conclusions.

\begin{figure*}[!htb]
	\centering\includegraphics[width=\textwidth]{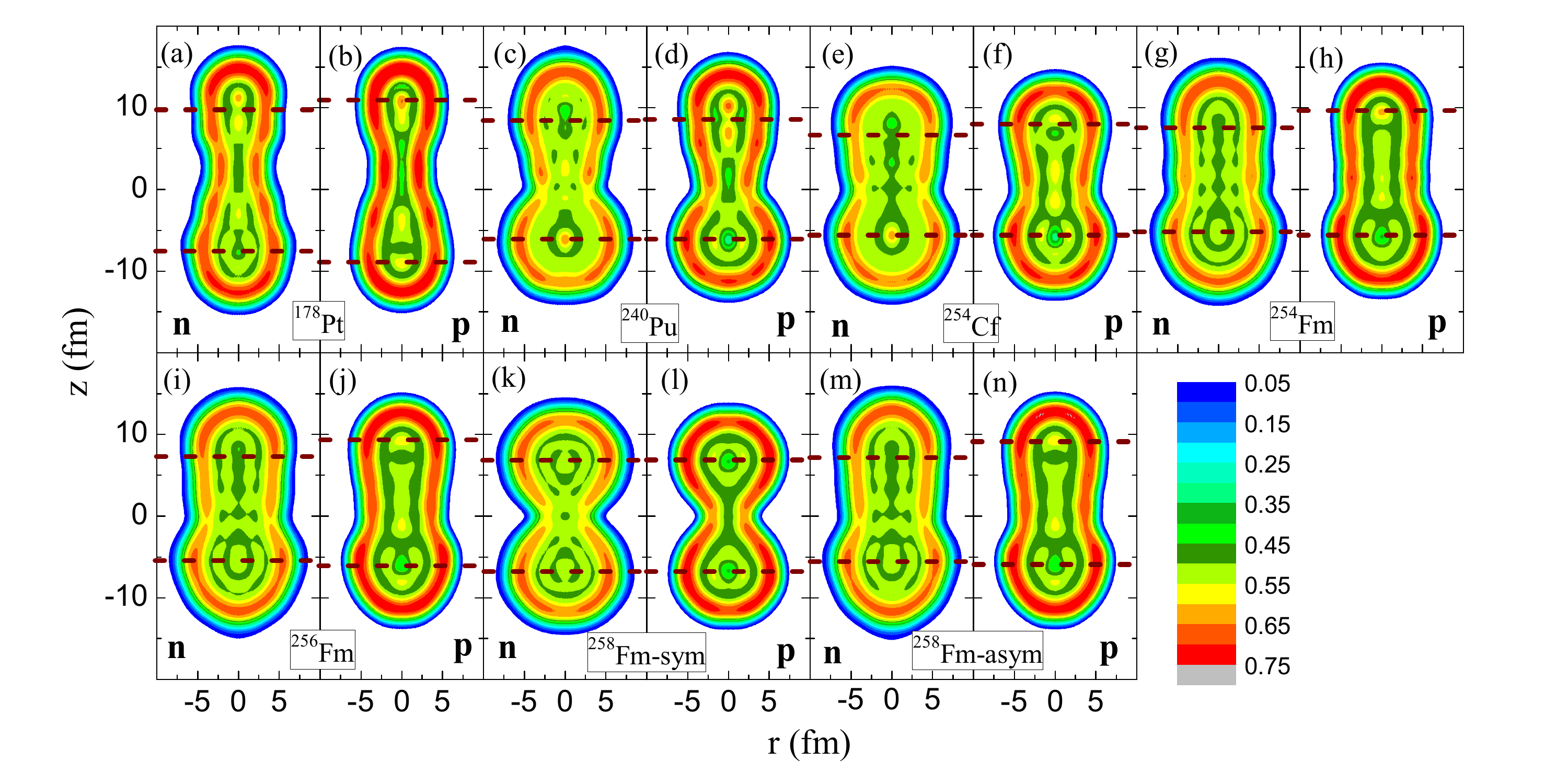}
	\caption{NLFs of selected nuclei used in the calculation of spontaneous
	fission fragment yield distributions. (a), (b), (g)-(n) are calculated with UNEDF1$_\text{HFB}$ EDF and (c)-(f) are calculated with SkM* EDF. Dashed lines indicate equatorial
	planes of the prefragments.}
	\label{fig:NLFprefrags}
\end{figure*}

\section{THEORETICAL FRAMEWORK}\label{Smodel}

\subsection{Selection of configurations to calculate NLFs}
Our method of estimating primary fission yields is based on NLFs computed at a
point on the most probable fission path that corresponds to a compact
configuration at the outer turning point well before scission. This
configuration, obtained by minimizing the collective action between the ground
state and the outer turning line, determines the peak position of the fragment
yield distribution. Specifically, for a given nucleus and fission decay
(spontaneous, neutron-induced or fusion-fission) we find the point where the
total energy drops down from the fission barrier to its ground state value
($E_\text{GS}$) along the most-probable fission path.  In case of SF, this point
lies on the outer turning line. For this configuration, called ${\mathcal C}$ in
the following, the two {\it prefragments} are identified using nucleonic
densities and NLFs. As discussed earlier, NLFs are related to the underlying
shell structure and remain almost unaffected along an effective fission
path~\cite{sadhukhan2017,Sim14,Mos71}.  If several fission pathways are
close in energy (multimodal fission) \cite{staszczak2009}, multiple nuclear
configurations may contribute to the yield distribution. For example, both
mass-symmetric and mass-asymmetric structures must be accounted for when the
corresponding fission valleys are close. 

The microscopic DFT input has been generated by using the HFB solver
HFODD~\cite{schunck2017}. In this study, we applied two Skyrme energy density
functionals (EDFs) in the particle-hole channel: SkM*~\cite{bartel1982} and
UNEDF1$_\text{HFB}$~\cite{schunck2015}. In the pairing channel, we took the
mixed-type density-dependent delta interaction~\cite{dobaczewski2002}. For
neutron-induced fission, the PES is obtained using the finite-temperature
Hartree-Fock-Bogoliubov (HFB) formalism, in which the compound nucleus is
described using the grand-canonical ensemble \cite{Pei09,martin2009,Sch15}.

\subsection{Definition of prefragments from NLF}
NLF measures the probability of 
finding two nucleons with the same spin $\sigma$ and
isospin $q$ at the same spatial localization. In this work, it is computed
as described in Refs.~\cite{reinhard2011,zhang2016}: 
\begin{equation}
	C_{q \sigma} = 
	\left[ 1 + 
	\left( 
	\frac{\tau_{q \sigma} \rho_{q \sigma} - \frac{1}{4} | \bm{\nabla} \rho_{q
	\sigma} |^{2} - \bm{j}^{2}_{q \sigma}}{\rho_{q\sigma}
	\tau^\text{TF}_{q \sigma}} 
	\right)^{2} 
	\right]^{-1} \,,
\end{equation}
where $\rho_{q\sigma}$, $\tau_{q\sigma}$, $\bm{j}_{q\sigma}$ and $\bm{\nabla}
\rho_{q\sigma}$ are the particle density, kinetic energy density, current
density, and density gradient, respectively. The Thomas-Fermi kinetic energy
density $\tau_{q\sigma}^\text{TF} = \frac{3}{5} (6 \pi^2)^{2/3}
\rho_{q\sigma}^{5/3}$ is introduced as a normalization parameter. 

Figure~\ref{fig:NLFprefrags} shows the NLFs used in the calculation of
spontaneous fission (SF) yields in the selected  nuclei. We recall that a value
of $C \sim 1$ indicates a large nucleon's localization, i.e., a low probability
of finding two nucleons with the same spin and isospin at the same spatial
localization. On the other hand, $C = 1/2$ corresponds to a limit of an
homogeneous Fermi gas. 
\begin{figure}[tbh]
	\includegraphics[width=1.0\columnwidth]{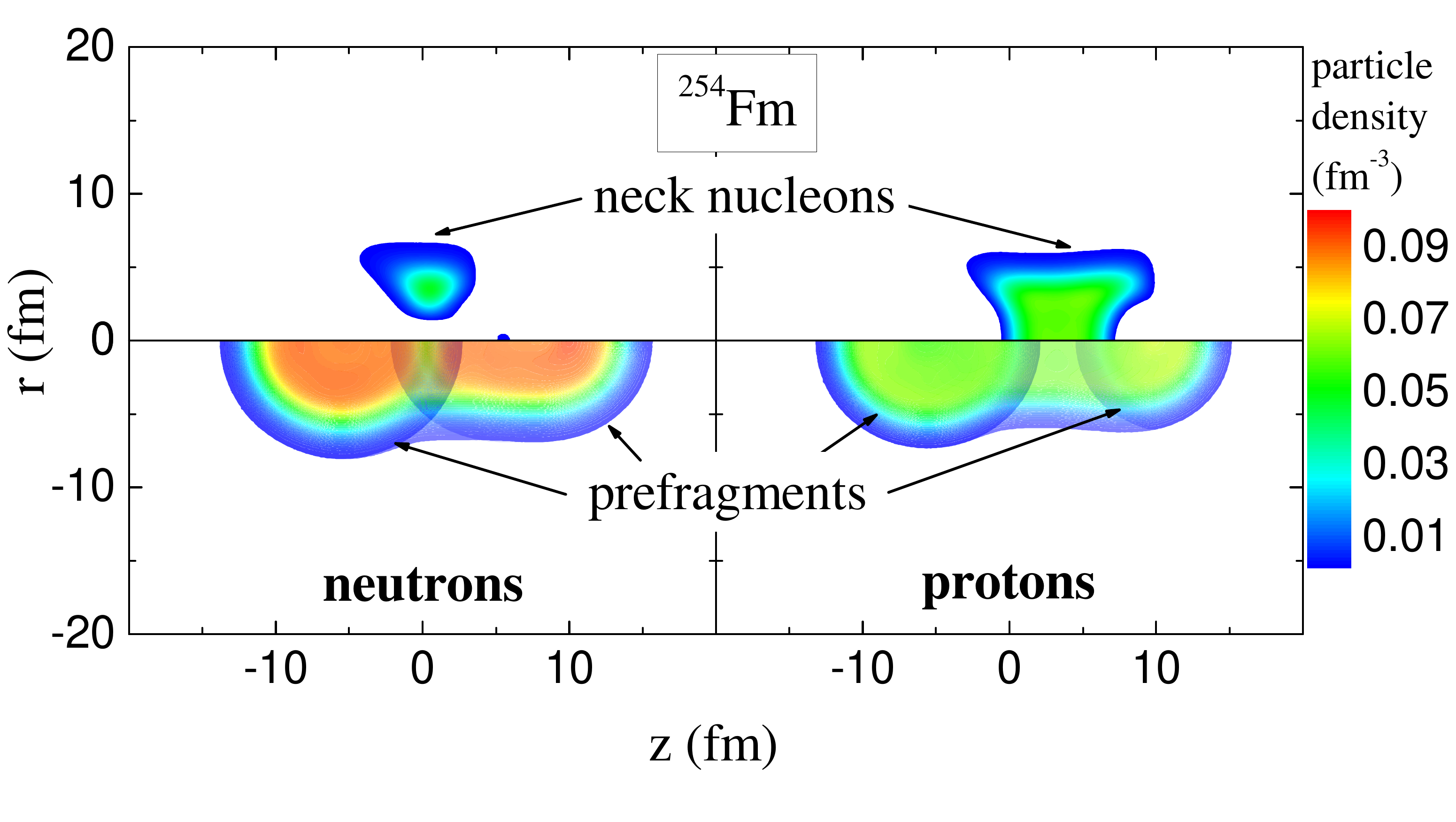}
	\caption{Density distributions of prefragments (bottom)  and neck-nucleons (top) in
	 the configuration $\cal C$ of $^{254}$Fm indicated in Fig.~\ref{fig:fm-pes}(a),  obtained  with  UNEDF1$_\text{HFB}$. \label{fig:neck}}
\end{figure}

The prefragments are determined using the following algorithm. Along the nuclear
symmetry axis, we identify the center of the prefragment as the NLF maximum (or
minimum) placed at the center of the NLF's concentric shell rings as shown in
Fig.~\ref{fig:NLFprefrags}.  In this way, the equatorial plane passing through
this center delimits the prefragment's hemisphere. Proton and neutron numbers of
prefragments are obtained by integrating the nucleon density over such
hemispheres and doubling the resulting number. This method is a generalization
of the previous definition~\cite{sadhukhan2017}.  As evident from
Fig.~\ref{fig:NLFprefrags} and discussed below, in most of fissioning actinide
nuclei, one of the prefragments is spherical due to the shell structure of the
doubly-magic nucleus $^{132}$Sn, while the other has an elongated shape with two
local NLF maxima separated by the central  minimum. 

\begin{figure}[tbh]
	\includegraphics[width=1.0\columnwidth]{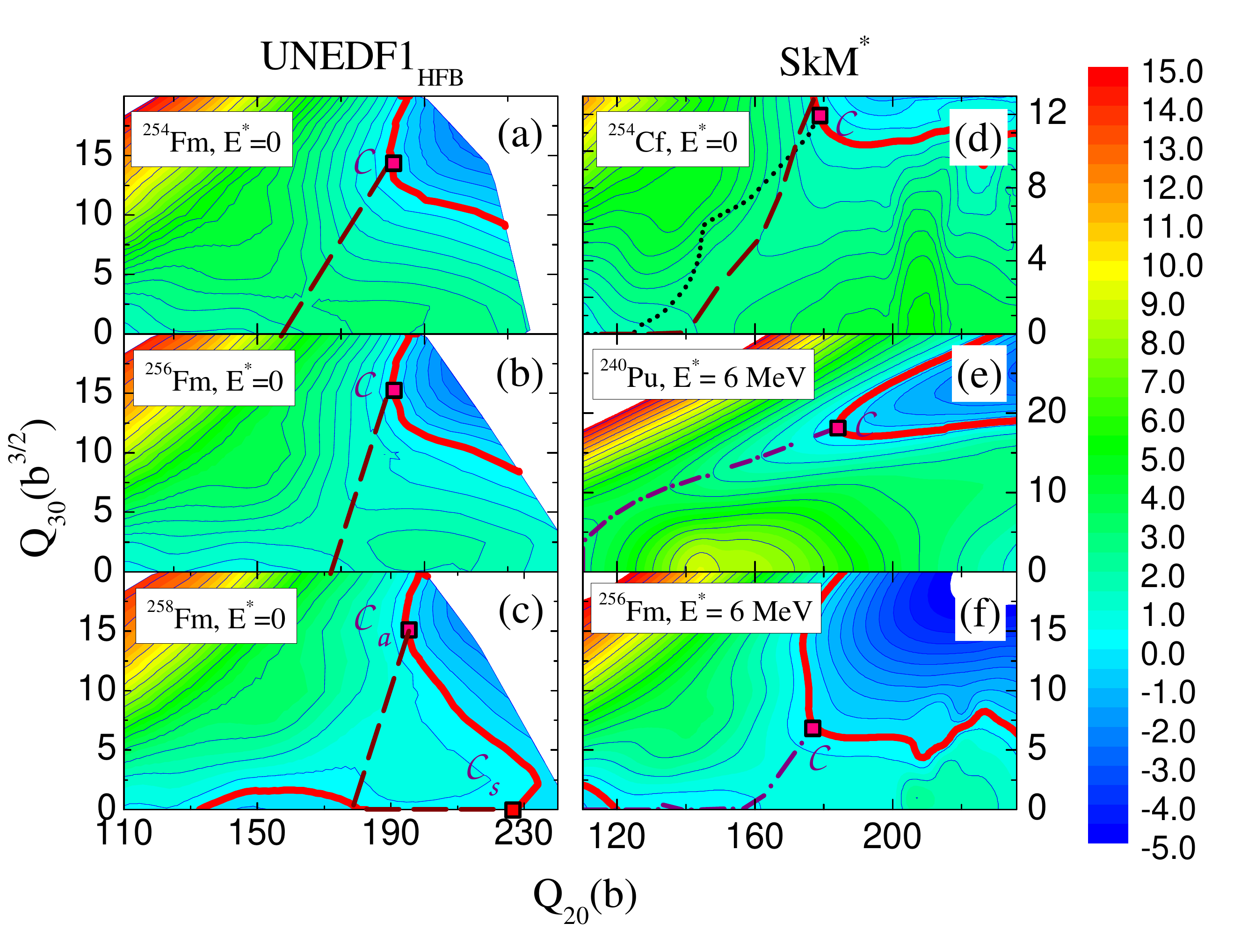}
	\caption{PESs of $^{254,256,258}$Fm, $^{254}$Cf, and $^{240}$Pu
		calculated with UNEDF1$_\text{HFB}$ (left panels) and
		SkM$^*$ (right panels).  For each nucleus, the energy (in
		MeV) is normalized to the ground state energy $E_\text{GS}$. The contours  $E
		= E_\text{GS}$  are indicated by thick solid lines. Most
		probable fission paths are calculated dynamically for $E^*=0$
		with constant inertia (dashed lines in (a)-(d)) and
		non-perturbative cranking inertia (dotted line in (d)) while static pathways are computed for $E^*>0$ (dash-dotted lines in (e) and
		(f)).  The configurations $\cal C$ used to compute prefragments
		are marked by squares. For $^{258}$Fm, two configurations are
		shown: ${\cal C}_s$ (symmetric fission pathway) and ${\cal C}_a$
		(asymmetric fission pathway).
		}
	\label{fig:fm-pes}
\end{figure}

Thus, a deformed fissioning nucleus can be thought of as two well-formed
prefragments connected by a ``glue'' of nucleons in the neck. This concept is
illustrated in Fig.~\ref{fig:neck}, which shows the neutron and proton
prefragments, and the neck nucleons for
	 the configuration $\cal C$ of $^{254}$Fm indicated in Fig.~\ref{fig:fm-pes}(a). 

\subsection{Yield distributions from prefragments}
 In the next step, the neck nucleons are distributed among the two prefragments
 with all possible combinations sampled. For each combination, binding energies
 of the resulting {\it fragments} ($E_{b1}$ for the fragment ($A_1,Z_1$) and
 $E_{b2}$ for the fragment ($A_2,Z_2$)) are calculated using the
 isospin-dependent liquid drop model~\cite{Mye66} (It is to be noted that the
 minimum mass and charge of a fragment is that of a corresponding prefragment). We again emphasize
 that the prefragment configurations are obtained using the most
 probable fission path calculated using realistic inputs. The liquid drop energy
 essentially determines the width of the yield distribution which primarily
 depends on the global isovector properties such as symmetry and Coulomb
 energies.

The energy of each fragment combination is $E_{r}=E_t-(E_{b1}+E_{b2}+E_{\rm
C})$, where $E_t$ is the calculated energy of the fissioning nucleus at
$\mathcal{C}$ (which is equal to $E_\text{GS}$) and $E_{\rm C}$ is the
electrostatic repulsion energy between the fragments. We assumed
the charges to be that of two point charges located at the center of mass of the
fragments. For completeness, we checked that the correction in the yield
distributions due to deformed charge distributions is negligible.
Subsequently, each pair of fragments is associated with a microcanonical
probability distribution~\cite{fong1953} 
\begin{eqnarray}\label{eq:prob}
	&P(A_1,A_2) \propto \sqrt{ \left(\frac{(A_1A_2)^8}{\left(A_1^{5/3}+A_2^{5/3}\right)^3(A_1+A_2)^3}\right) \frac{a_1 a_2}{(a_1+a_2)^5} 
	}
	\nonumber\\
	&\times\left(1-\frac{1}{2\sqrt{(a_1+a_2)E_r}}\right)
	E_r^{9/4}\exp{\left\{2\sqrt{(a_1+a_2)E_r}\right\}},
\end{eqnarray}
where $a_i=A_i/10$ MeV$^{-1}$ is the level density
parameter~\cite{sadhukhan2016}. Finally, we fold the probabilities with a
Gaussian smoothing function of width 3 for $A$ and 2 for $Z$
\cite{sadhukhan2016}.

To asses the robustness of our calculations, we estimated uncertainties in
the yield distributions with respect to the model parameters, including the choice of the EDF. The distributions
are found to be very sensitive to the prefragment particle numbers. In general, the neutron and proton numbers of a prefragment are not integers as they are obtained by integrating density distributions. Hence, we consider the nearest integer as average particle (neutron or proton) number with a spread of $\pm$1 around this average value. This particle number uncertainty is considered for both the prefragments of a fissioning system. The  yield distributions are then obtained for all possible combinations among these particle numbers. The resulting uncertainty in the yield distributions is  referred to as  the two-particle uncertainty. On the other hand, uncertainties from different realizations of
the liquid drop model and other parameters defining the microcanonical
probability mainly modify the tail part of the distributions and its magnitude
is smaller than the prefragment-uncertainty.   Also, we checked that the absolute
value of $E_r$ weakly affects yield distributions. We conclude therefore
that, while a more microscopic treatment of this quantity could be important for
estimating observables such as the total kinetic energy of the fragments, its
impact on the fragments yields is minor within our model. 

\section{RESULTS}\label{Sresults}
\subsection{Benchmarking the model against experimental data}
To benchmark our model, we consider several nuclei ($^{178}$Pt, $^{240}$Pu,
$^{254}$Cf, and $^{254,256,258}$Fm) for which  fission yield distributions have
been determined experimentally. For each of these systems, an appropriate
configuration $\cal C$ is identified to extract the prefragments.  In case of
thermal fission $(n_{th},f)$, we calculate the PES at the thermal excitation
energy ($\approx 6$ MeV) and we select the configuration $\cal C$ by taking the
point on the static fission path that has zero energy with respect to the ground
state minimum obtained at $E^* = 6$~MeV. For pre-actinide nuclei such as
$^{178}$Pt, the potential barrier is fairly flat and broad, and it is crossed by
the fusion valley~\cite{warda2012a,tsekhanovich2019}. Therefore, we have
arbitrarily chosen a configuration on the static paths of
$^{178}$Pt~\cite{tsekhanovich2019} at $Q_{20}=220$\,b (UNEDF1$_\text{HFB}$) and
190\,b (SkM*). We checked that the prefragment structure does not change beyond
these points.

The isotopes of Fm are of special interest as the corresponding mass yields show
a transition from asymmetric to a symmetric distribution with increasing parent
mass \cite{hulet1989,moller1989,Brosa90,staszczak2009,regnier2019}. Calculated
PESs are shown in Fig.~\ref{fig:fm-pes}. For all the systems except $^{258}$Fm,
fission fragment distributions are expected to be asymmetric because of a large
outer fission barrier. However, the potential surface of $^{258}$Fm is rather
flat at large $Q_{20}$ and we expect contributions from both symmetric and
asymmetric fission pathways. 

From previous calculations~\cite{Sadhukhan2014,zhao2015,sadhukhan2016}, it is
found that the topology of the minimum action path is rather simple on the
$Q_{20}$-$Q_{30}$ plane irrespective of the choice of collective inertia.
Usually, it follows the minimum distance from a mass-symmetric configuration
outside the fission isomer to the nearest outer turning point. This is verified
in Fig.~\ref{fig:fm-pes} where  the dynamical paths calculated with constant
inertia and non-perturbative cranking inertia~\cite{Baran2011,giuliani2018b}
are compared for $^{254}$Cf. Although there is a local detour inside the barrier due to
non-perturbative inertia, both the pathways reach the outer turning line at
nearby points at which the NLFs are practically identical. A similar behavior has been reported for the fission paths of $^{240}$Pu~\cite{sadhukhan2016}.
Therefore, as we are only concerned with identifying the appropriate $\cal C$
configuration, we assumed a constant inertia for the calculation of minimum
action paths of Fm isotopes and the relative weight of the two-configurations of
$^{258}$Fm. This reduces the computational burden enormously compared to
calculations with non-perturbative collective inertia.
\begin{figure}[htb]
	\centering\includegraphics[width=0.8\columnwidth]{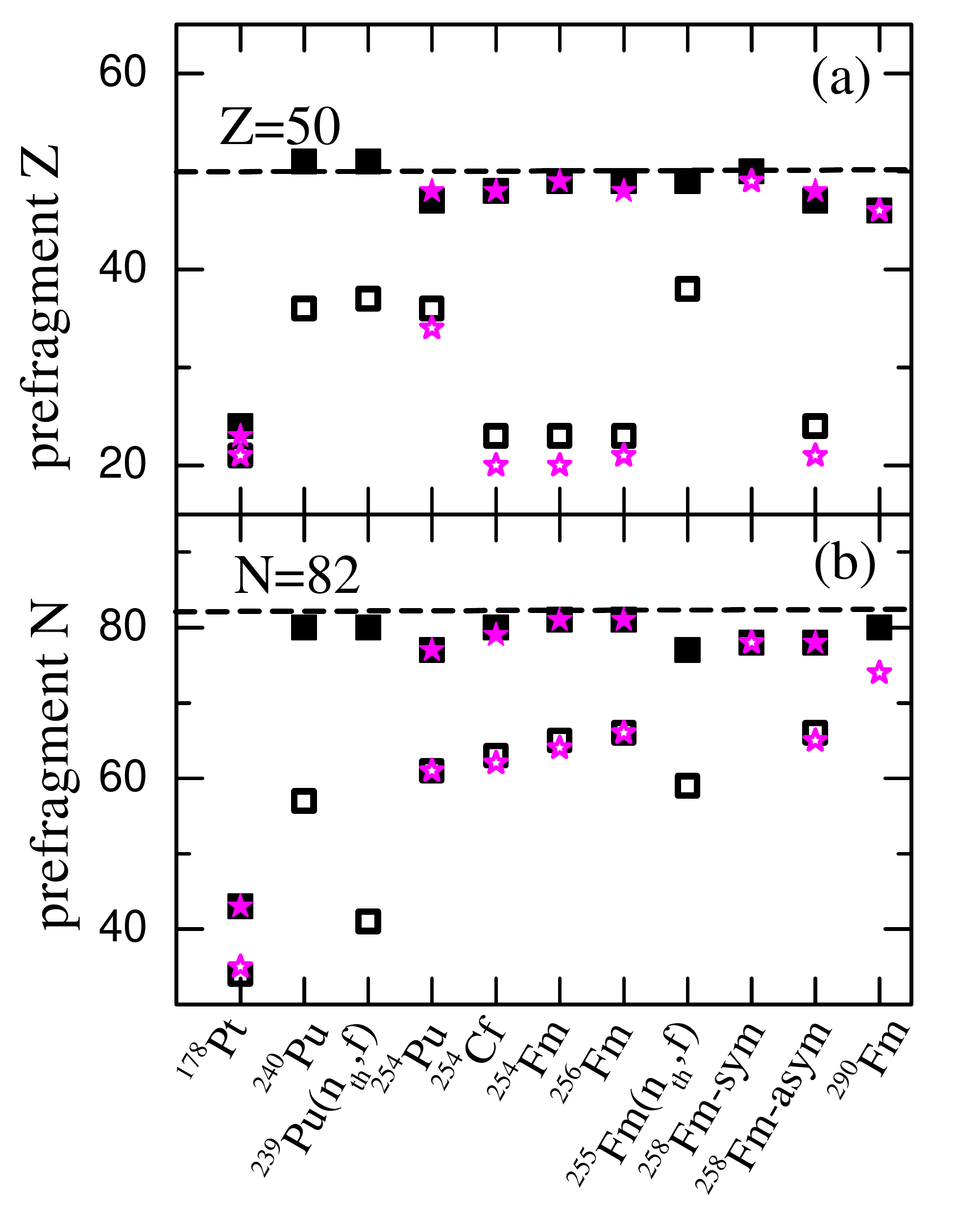}
	\caption{Prefragments for different fissioning systems calculated with
	SkM* (squares) and UNEDF1$_\text{HFB}$ (stars). Top: proton
	number; bottom: neutron number. Heavy and light prefragments are marked
	with closed and open symbols, respectively. Note that for  ${\cal C}_s$
	configurations of $^{258}$Fm and $^{290}$Fm only symmetric fission
	is expected and  hence only one symbol is shown.
	}
	\label{fig:prefrags}
\end{figure}

\begin{figure}[tbh]
	\includegraphics[width=0.7\columnwidth]{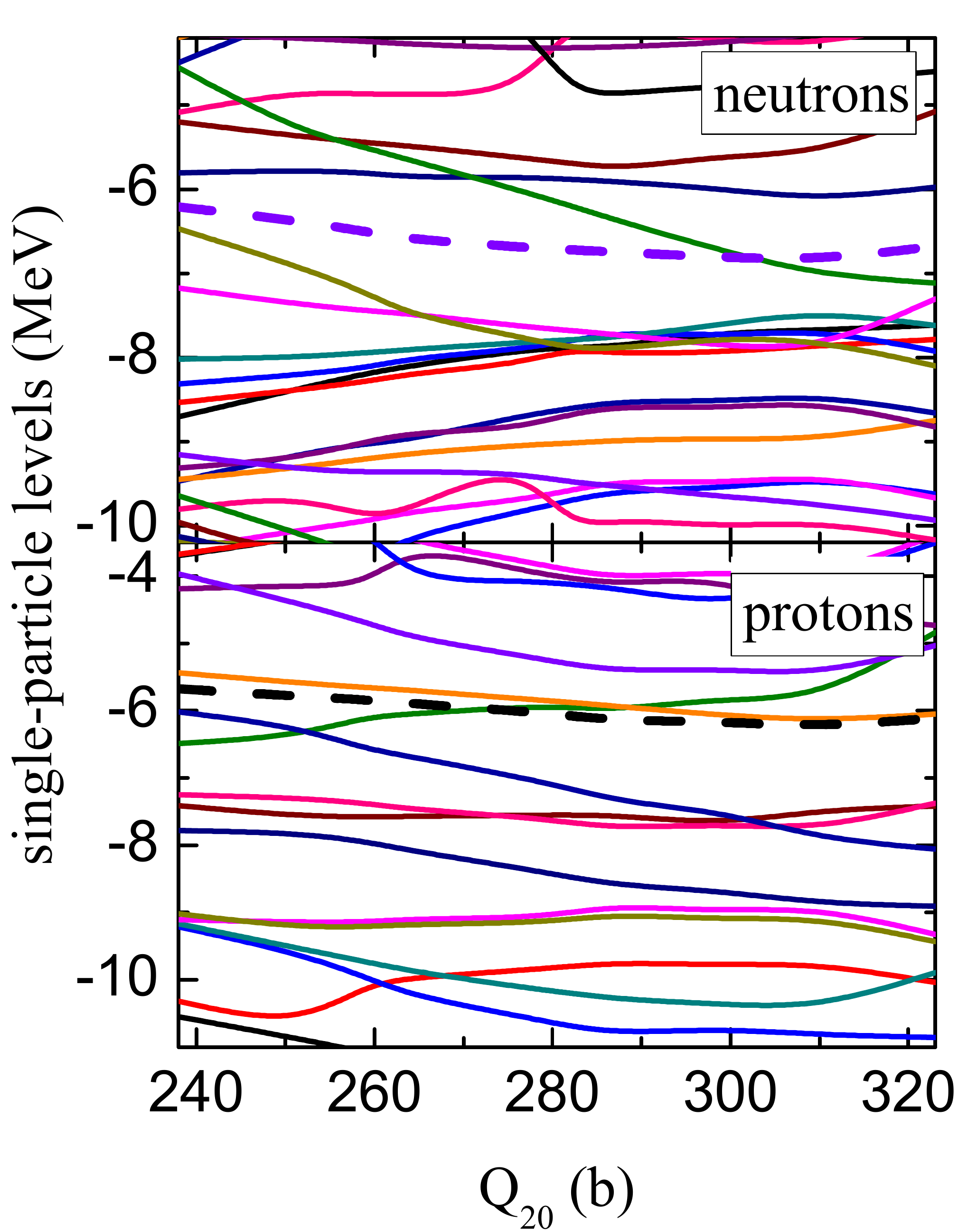}
	\caption{Single particle levels calculated with SkM$^*$ along the most
	probable fission path of $^{240}$Pu from the outer turning point to
	scission. Fermi energies are indicated by thick dashed lines.}
	\label{fig:240Pu-levels}
\end{figure}

To complement the NLFs  shown in Fig.~\ref{fig:NLFprefrags},
Fig.~\ref{fig:prefrags} displays the calculated prefragments for different
fissioning systems.  We notice that for the actinides the heavy prefragment is
very close to the doubly-magic nucleus $^{132}$Sn irrespective of the
EDF used. This result suggests an early development of shell effects
along the fission path, in agreement with the previous NLF-based study on the
formation of $^{240}$Pu fission fragments~\cite{sadhukhan2017}. On the other
hand, both mass and charge of the lighter fragment are subject to appreciable
fluctuations; this supports the need for the microscopic determination of
prefragments. As already pointed out in Ref.~\cite{sadhukhan2017}, the main
advantage of using NLFs is that the localization patterns of the prefragments
defined at the outer turning line closely resemble those at scission. This can
be confirmed by the fact that both proton and neutron single particle levels
show a smooth pattern in their evolution from the outer turning point to the
scission point, indicating the stabilization of shell
effects~\cite{negele1989,nazarewicz1993,sadhukhan2013,scamps2018a}. In
Fig.~\ref{fig:240Pu-levels}, we show the single-particle levels of $^{240}$Pu
along the most-probable effective fission path of Ref.~\cite{sadhukhan2017}.
This plot shows that both neutron and proton energy levels change very gradually
maintaining the low single-particle level density around the Fermi level. An
additional factor that maintains the adiabacity  of the collective motion is the
presence of pairing correlations, which reduce the effect of configuration
changes~\cite{(mor74),negele1989,nazarewicz1993,Sadhukhan2014,Bernard2019}.  

\begin{figure}[htb]
	\includegraphics[width=1.0\columnwidth]{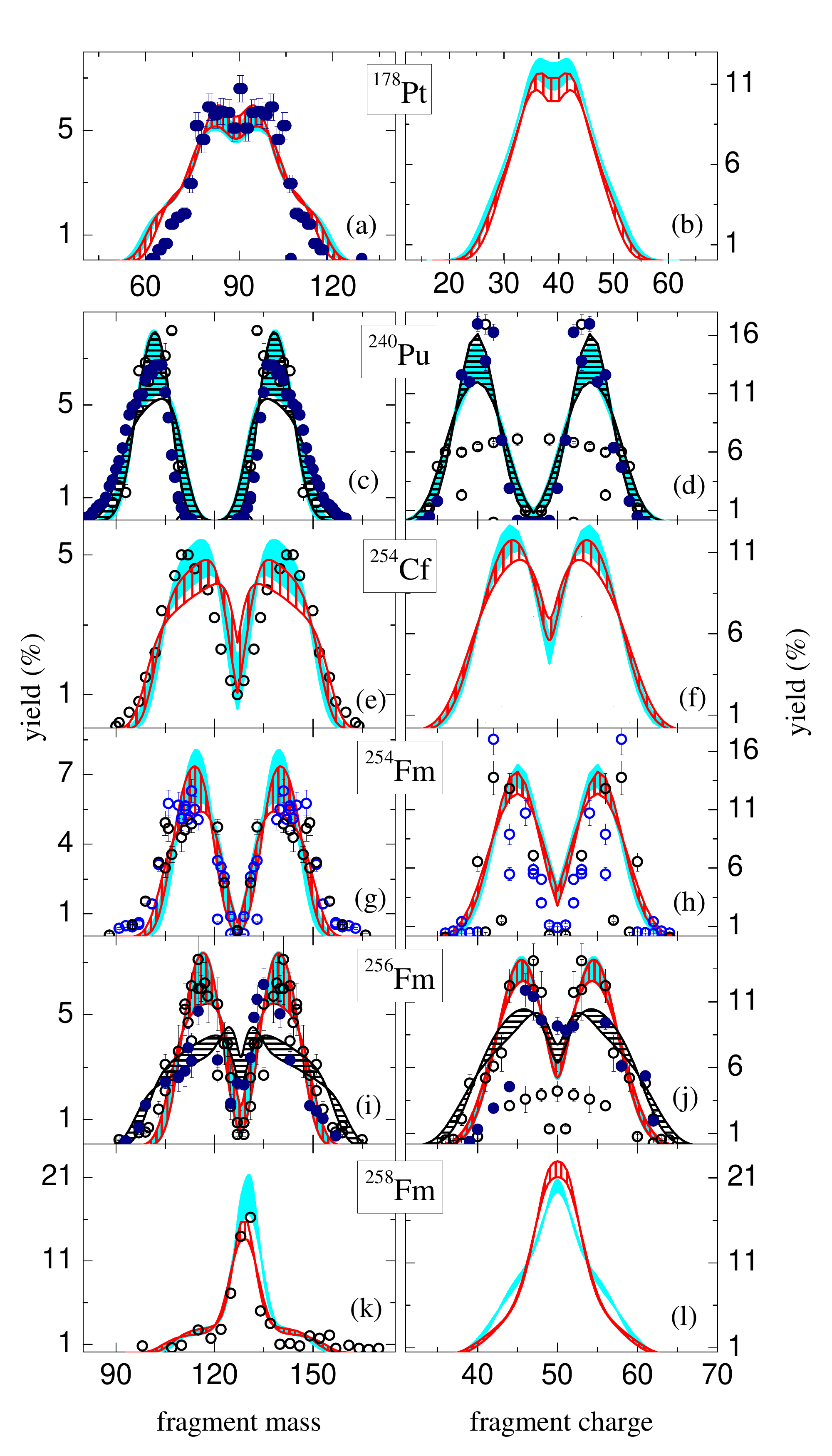}
	\caption{Calculated yield distributions for SF (shaded for SkM* and
		vertically patterned for UNEDF1$_\text{HFB}$) and thermal
		fission (horizontally patterned) with two-particle uncertainty
		in each prefragment.  Open circles: experimental
		data for SF of $^{240}$Pu~\cite{Laidler1962},
		$^{254}$Cf~\cite{brandt1963},
		$^{254}$Fm~\cite{gindler1977,harbour1973},
		$^{256}$Fm~\cite{flynn1972},
		$^{258}$Fm~\cite{hoffman1980}; closed circles:
		experimental data for thermal fission of
		$^{240}$Pu~\cite{schmitt1984} and
		$^{256}$Fm~\cite{Fly75}, and heavy-ion induced fission of
		$^{178}$Pt~\cite{tsekhanovich2019}. 
	}
	\label{fig:exp-yields}
\end{figure}

Since the notion of a prefragment is a purely theoretical concept, their
properties cannot be measured experimentally. Moreover, alternate prefragment
definitions exist~\cite{scamps2018a,scamps2019a}. Consequently, the validity of
a prefragment-based description is checked by the ability of a model to predict
and reproduce experimental observables. To this end, in
Fig.~\ref{fig:exp-yields}, we benchmark our approach by predicting the fission
fragment mass and charge distributions for selected nuclei. In general, very
good agreement with experiment has been obtained for SF. For thermal fission,
one does not expect a drastic change compared to SF as the excitation energy
is quite low.
Indeed, for $^{239}$Pu$(n_{th},f)$ both experiment and predictions overlap with
the SF fragment yields. For $^{255}$Fm$(n_{th},f)$ we predict broader mass and
charge distributions than for SF\@. We note that the $^{255}$Fm$(n_{th},f)$
experimental data show an asymmetry in the heavy and light fragment yields,
which may be attributed to neutron evaporations from the primary fragments. In
the case of $^{258}$Fm, the mass-symmetric trajectory produces a sharp peak in
the SF mass distribution while the asymmetric one contributes to the broad
tail, in agreement with the experimental mass distribution. Based on these
benchmark calculations, we conclude that the proposed method is robust
with respect to the choice of the EDF as the SkM* and UNEDF1$_\text{HFB}$ results are very close. The
information contained in the outer-point configurations $\cal C$, supplemented
by the statistical treatment of the neck nucleons is sufficient to predict the
mass and charge flows, which are governed by macroscopic liquid drop forces. 

\begin{figure}[tbh]
	\includegraphics[width=1.0\columnwidth]{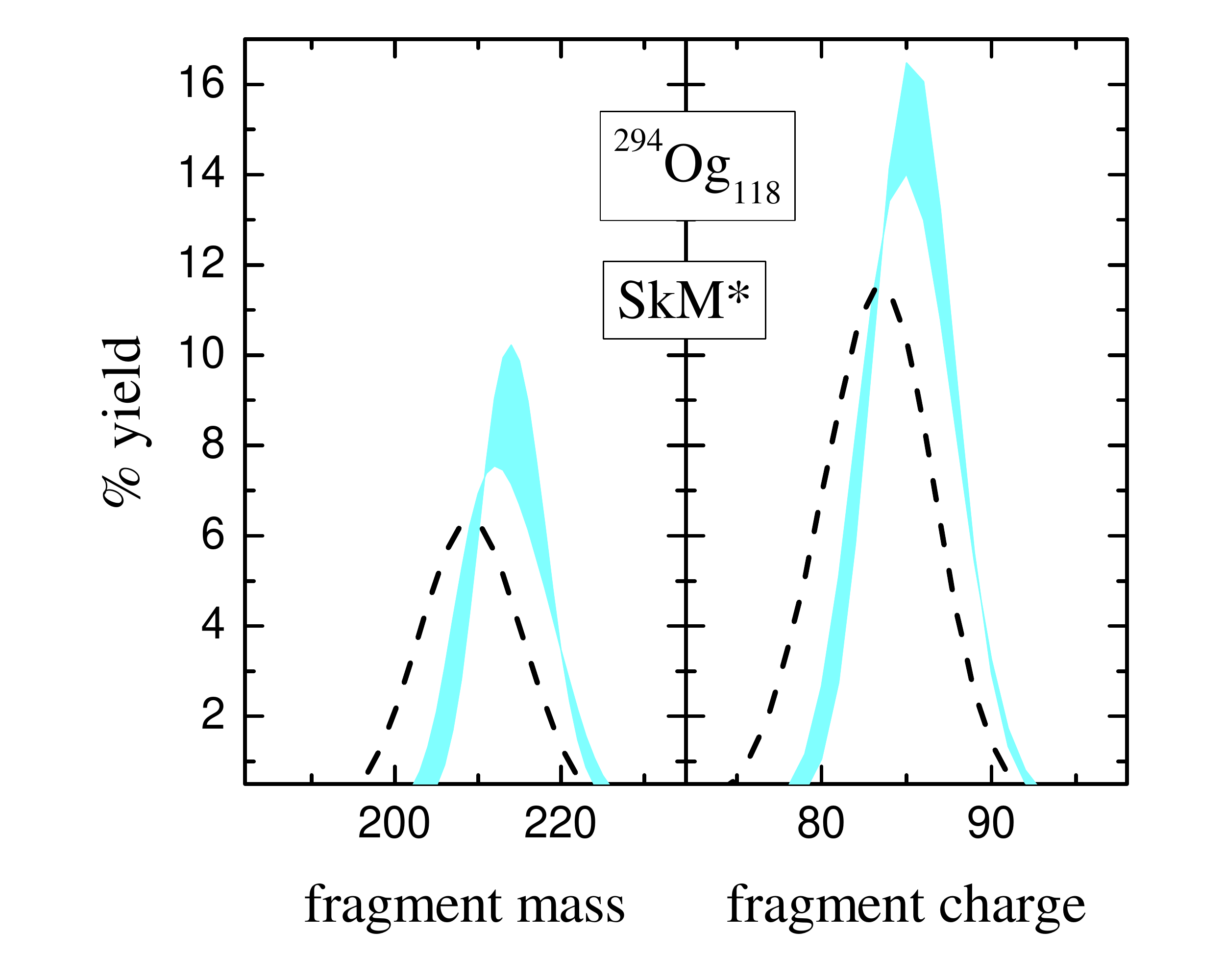}
	\caption{Yield distributions for $^{294}$Og calculated in SkM$^*$ using the statistical model described in this work  (shaded
	region) and the Langevin approach of Ref.~\cite{matheson2019} (dashed lines). Only heavy-fragment distributions are plotted.}
	\label{fig:294Og-yields}
\end{figure}

\subsection{Comparison with Langevin results for $^{294}$Og}
Our approach to fission fragments distributions is rooted in the  microscopic Langevin 
calculations suggesting a separation of scales between the formation of the
prefragments and the rapid rearrangement of the neck nucleons at scission~\cite{sadhukhan2016}. To further prove the capability of our model to
retain the relevant physics captured by Langevin calculations,
Fig.~\ref{fig:294Og-yields} compares the  fission fragments distributions for $^{294}$Og
predicted in our model with those predicted
 in Ref.~\cite{matheson2019} using the Langevin approach. Both models are in a fairly good agreement,
predicting the emergence of cluster decay (with the heavy fragment centered
around $^{208}$Pb) as a main fission mode. This again confirms the ability of the present model to capture the
relevant physics. We notice that our calculations
predict a slight shift towards a more asymmetric split, and a narrower
distribution of yields. Both effects can be related to the lower mass and charge limits
imposed by the identification of the prefragments, which drive the final
configurations towards a heavier fragment and slightly reduce the
fragment phase space. 

\subsection{Application to \rpa\ nuclei}
\begin{figure}[tb]
	\includegraphics[width=1.0\columnwidth]{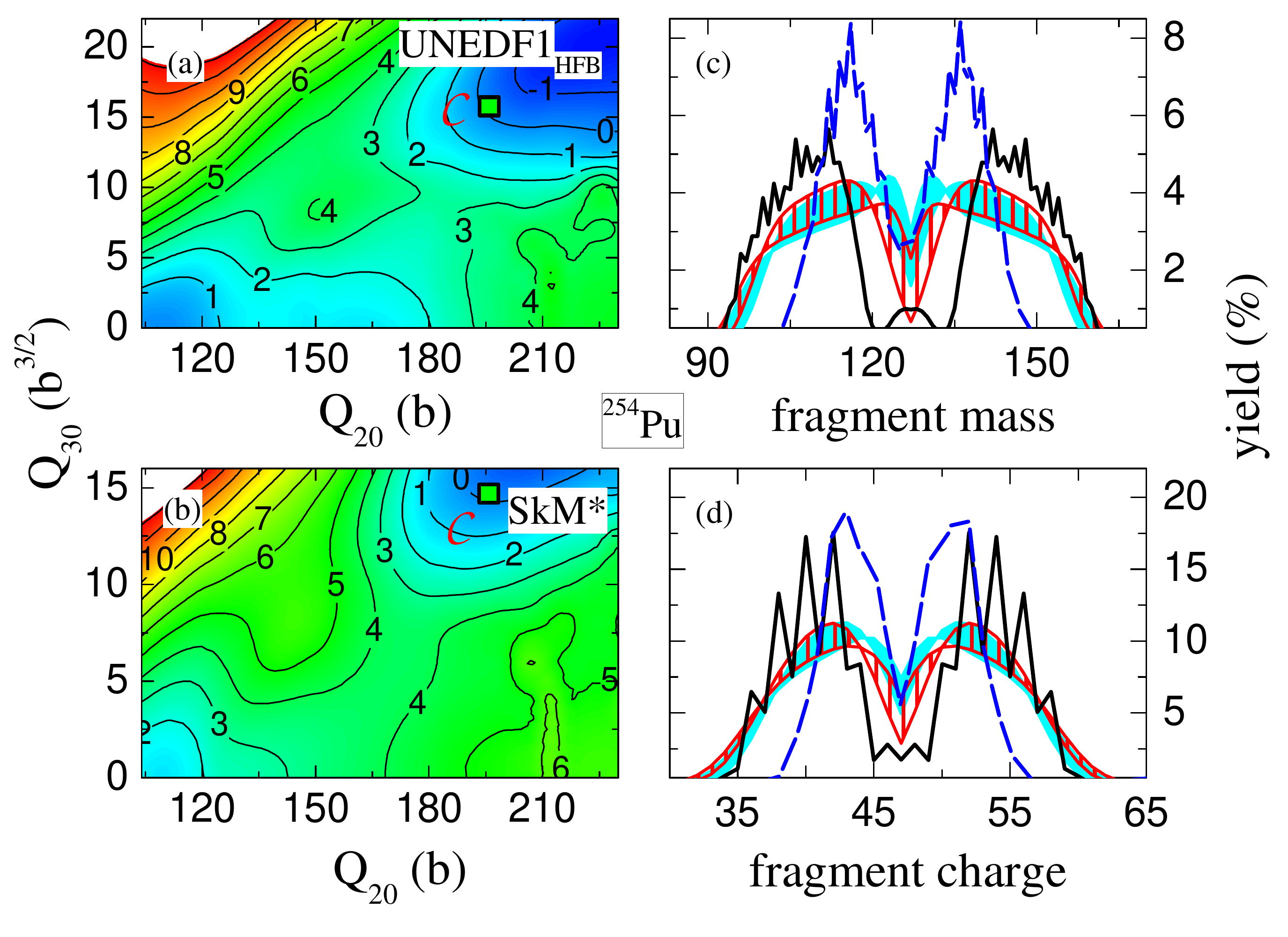}
	\caption{Predicted fission properties of {\rpa} nucleus $^{254}$Pu.
	 Left: Potential energy surfaces calculated with (a) UNEDF1$_\text{HFB}$ and (b) SkM* functionals. The configurations
	${\cal C}$ are marked by square symbols. Right: Predicted (c) mass and (d) charge
	yield distributions (shaded and patterned regions for SkM* and UNEDF1$_\text{HFB}$, respectively) given by our method. The
	GEF~\cite{schmidt2016} and ABLA~\cite{kelic2009}
	results are shown by solid and dashed lines, respectively.}
	\label{fig:rproc-yields1}
\end{figure}

\begin{figure}[tb]
	\includegraphics[width=1.0\columnwidth]{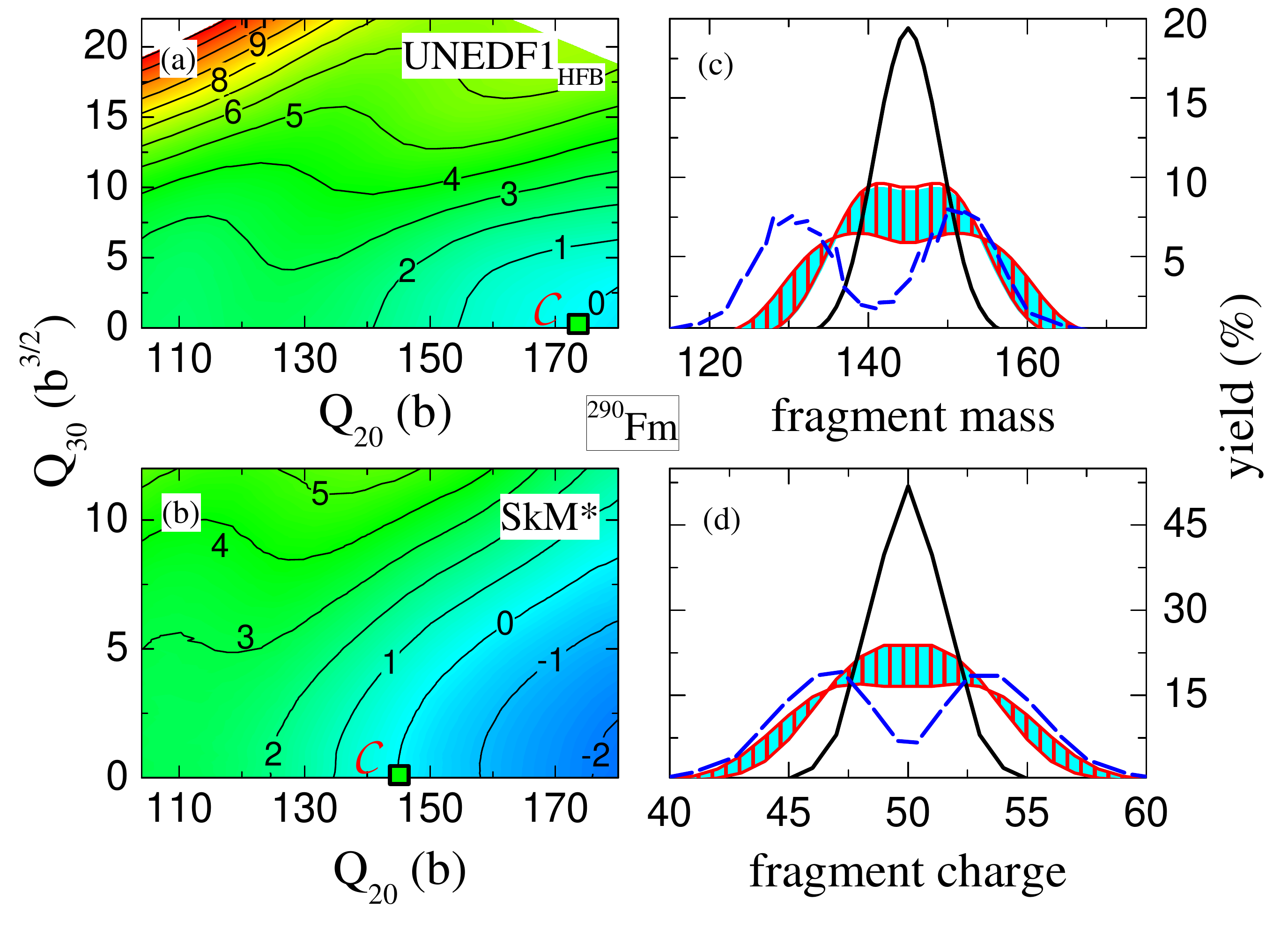}
	\caption{Similar as in  Fig. \ref{fig:rproc-yields1} but for $^{290}$Fm.
	The predictions of UNEDF1$_\text{HFB}$ and  SkM* are practically identical.}
	\label{fig:rproc-yields2}
\end{figure}

Encouraged by the positive outcome of the benchmarking exercise, we extended our
calculations to {\rpa} nuclei. To this end, we consider $^{254}$Pu as a
representative fissioning nucleus that is expected to significantly contribute
to the \rpa\ nucleosynthesis occurring in neutron star mergers
 and the extremely neutron-rich
$^{290}$Fm nucleus which is speculated to be formed at the edge of
{\rpa}~\cite{vassh2019}.  A one-dimensional fission trajectory is calculated for
each isotope up to the fission isomer by constraining $Q_{20}$ and leaving the
other degrees of freedom unconstrained. In the region between the fission isomer
and the outer turning line, the collective space is expanded to include $Q_{30}$
in order to account for mass asymmetry. The resulting PESs and yield
distributions are shown in Fig.~\ref{fig:rproc-yields1} and
Fig.~\ref{fig:rproc-yields2} for $^{254}$Pu and $^{290}$Fm, respectively. These
are compared with predictions given by the semi-empirical GEF~\cite{schmidt2016}
and ABLA~\cite{kelic2009} models.

For $^{254}$Pu, our method predicts yield distributions somewhere in between the
GEF and ABLA results. Distributions are wider as predicted by GEF and also have
considerable symmetric contributions as in ABLA\@. For $^{290}$Fm, our approach
predicts a symmetric fission centered around $^{145}$Sn, but with a wider
fragment distribution compared to GEF and closer to the width predicted by the
ABLA model. This result suggests a weakening of $Z/N=50/82$ shell stabilization
for very neutron-rich nuclei, in agreement with a general trend found in recent
calculations~\cite{mumpower2019} that would extend the range of \rpa\ elements
affected by fission cycling~\cite{vassh2019a}.  It is worth noting that
while in both GEF and ABLA the effect of shell stabilization determining the
fission modes is introduced \textit{ad hoc} using parameters finely tuned to
reproduce experimental data, in our approach they naturally emerge from the
underlying EDF\@. This is a crucial feature for the studies of exotic rare
isotopes whose shell closures may differ from those found in nuclei close to the
line of beta stability. Finally, we point out the extreme robustness of the
fission yields with respect to the choice of EDF even in this region far from stability.
This feature, which has been also found in our DFT+Langevin
studies~\cite{matheson2019}, gives us the confidence to extend our study to systematic calculations or \rpa\ nuclei.

\section{CONCLUSIONS}\label{Sconclusions} 
We developed a method for estimating fission fragment
yields based on nuclear DFT with realistic energy density functionals and simple
statistical assumptions governing the redistribution of neck nucleons at
scission. This method is inspired by recent microscopic Langevin calculations
suggesting an early formation of the prefragments along effective fission paths
and a rapid distribution of neck nucleons around the scission
point~\cite{sadhukhan2017}, which is reminiscent of the separability principle
proposed in previous phenomenological studies~\cite{schmidt2008}. The new
approach is  computationally inexpensive compared to more microscopic models
that require the knowledge of the full multi-dimensional PES and related
quantities (such as the collective inertia) all the way to scission, while
retaining the relevant physics determining the  structure
of fission fragment distributions. A sound agreement with experimental charge and 
mass distributions is obtained for a wide range of fissioning nuclei.

In the next step, we intend to use the new method to carry out global
calculations of fission fragment distributions across the {\rpa} region. Such
input could be combined with systematic calculations of fission
rates~\cite{goriely2009,erler2012,giuliani2018a,giuliani2019a} to consistently
study the impact of fission on the \rpa\ nucleosynthesis. Moreover, the present formulation of the model, based on DFT computation of densities in even-even nuclei, is not capable of reproducing the even-odd effects of fission fragment yields \cite{Gonnenwein2013}. We will consider this aspect in our future studies.

\begin{acknowledgments}
	The authors are grateful to G.~Mart{\'i}nez~Pinedo and N.~Vassh for
	providing the ABLA and GEF data, respectively, used in
	Figs.~\ref{fig:rproc-yields1} and \ref{fig:rproc-yields2}.  This work was supported by the U.S.
	Department of Energy under Award Numbers DE-NA0003885 (NNSA, the
	Stewardship Science Academic Alliances program), DE-SC0013365 (Office of
	Science), and DE- SC0018083 (Office of Science, NUCLEI SciDAC-4
	collaboration). Computing support came from the Lawrence Livermore
	National Laboratory (LLNL) Institutional Computing Grand Challenge
	program.
\end{acknowledgments}

\bibliography{ref,books,main}

\begin{thebibliography}{81}%
\makeatletter
\providecommand \@ifxundefined [1]{%
 \@ifx{#1\undefined}
}%
\providecommand \@ifnum [1]{%
 \ifnum #1\expandafter \@firstoftwo
 \else \expandafter \@secondoftwo
 \fi
}%
\providecommand \@ifx [1]{%
 \ifx #1\expandafter \@firstoftwo
 \else \expandafter \@secondoftwo
 \fi
}%
\providecommand \natexlab [1]{#1}%
\providecommand \enquote  [1]{``#1''}%
\providecommand \bibnamefont  [1]{#1}%
\providecommand \bibfnamefont [1]{#1}%
\providecommand \citenamefont [1]{#1}%
\providecommand \href@noop [0]{\@secondoftwo}%
\providecommand \href [0]{\begingroup \@sanitize@url \@href}%
\providecommand \@href[1]{\@@startlink{#1}\@@href}%
\providecommand \@@href[1]{\endgroup#1\@@endlink}%
\providecommand \@sanitize@url [0]{\catcode `\\12\catcode `\$12\catcode
  `\&12\catcode `\#12\catcode `\^12\catcode `\_12\catcode `\%12\relax}%
\providecommand \@@startlink[1]{}%
\providecommand \@@endlink[0]{}%
\providecommand \url  [0]{\begingroup\@sanitize@url \@url }%
\providecommand \@url [1]{\endgroup\@href {#1}{\urlprefix }}%
\providecommand \urlprefix  [0]{URL }%
\providecommand \Eprint [0]{\href }%
\providecommand \doibase [0]{http://dx.doi.org/}%
\providecommand \selectlanguage [0]{\@gobble}%
\providecommand \bibinfo  [0]{\@secondoftwo}%
\providecommand \bibfield  [0]{\@secondoftwo}%
\providecommand \translation [1]{[#1]}%
\providecommand \BibitemOpen [0]{}%
\providecommand \bibitemStop [0]{}%
\providecommand \bibitemNoStop [0]{.\EOS\space}%
\providecommand \EOS [0]{\spacefactor3000\relax}%
\providecommand \BibitemShut  [1]{\csname bibitem#1\endcsname}%
\let\auto@bib@innerbib\@empty
\bibitem [{\citenamefont {Horowitz}\ \emph {et~al.}(2019)\citenamefont
  {Horowitz}, \citenamefont {Arcones}, \citenamefont {C{\^{o}}t{\'{e}}},
  \citenamefont {Dillmann}, \citenamefont {Nazarewicz}, \citenamefont
  {Roederer}, \citenamefont {Schatz}, \citenamefont {Aprahamian}, \citenamefont
  {Atanasov}, \citenamefont {Bauswein}, \citenamefont {Beers}, \citenamefont
  {Bliss}, \citenamefont {Brodeur}, \citenamefont {Clark}, \citenamefont
  {Frebel}, \citenamefont {Foucart}, \citenamefont {Hansen}, \citenamefont
  {Just}, \citenamefont {Kankainen}, \citenamefont {McLaughlin}, \citenamefont
  {Kelly}, \citenamefont {Liddick}, \citenamefont {Lee}, \citenamefont
  {Lippuner}, \citenamefont {Martin}, \citenamefont {Mendoza-Temis},
  \citenamefont {Metzger}, \citenamefont {Mumpower}, \citenamefont
  {Perdikakis}, \citenamefont {Pereira}, \citenamefont {O'Shea}, \citenamefont
  {Reifarth}, \citenamefont {Rogers}, \citenamefont {Siegel}, \citenamefont
  {Spyrou}, \citenamefont {Surman}, \citenamefont {Tang}, \citenamefont
  {Uesaka},\ and\ \citenamefont {Wang}}]{horowitz2018}%
  \BibitemOpen
  \bibfield  {author} {\bibinfo {author} {\bibfnamefont {C.~J.}\ \bibnamefont
  {Horowitz}}, \bibinfo {author} {\bibfnamefont {A.}~\bibnamefont {Arcones}},
  \bibinfo {author} {\bibfnamefont {B.}~\bibnamefont {C{\^{o}}t{\'{e}}}},
  \bibinfo {author} {\bibfnamefont {I.}~\bibnamefont {Dillmann}}, \bibinfo
  {author} {\bibfnamefont {W.}~\bibnamefont {Nazarewicz}}, \bibinfo {author}
  {\bibfnamefont {I.~U.}\ \bibnamefont {Roederer}}, \bibinfo {author}
  {\bibfnamefont {H.}~\bibnamefont {Schatz}}, \bibinfo {author} {\bibfnamefont
  {A.}~\bibnamefont {Aprahamian}}, \bibinfo {author} {\bibfnamefont
  {D.}~\bibnamefont {Atanasov}}, \bibinfo {author} {\bibfnamefont
  {A.}~\bibnamefont {Bauswein}}, \bibinfo {author} {\bibfnamefont {T.~C.}\
  \bibnamefont {Beers}}, \bibinfo {author} {\bibfnamefont {J.}~\bibnamefont
  {Bliss}}, \bibinfo {author} {\bibfnamefont {M.}~\bibnamefont {Brodeur}},
  \bibinfo {author} {\bibfnamefont {J.~A.}\ \bibnamefont {Clark}}, \bibinfo
  {author} {\bibfnamefont {A.}~\bibnamefont {Frebel}}, \bibinfo {author}
  {\bibfnamefont {F.}~\bibnamefont {Foucart}}, \bibinfo {author} {\bibfnamefont
  {C.~J.}\ \bibnamefont {Hansen}}, \bibinfo {author} {\bibfnamefont
  {O.}~\bibnamefont {Just}}, \bibinfo {author} {\bibfnamefont {A.}~\bibnamefont
  {Kankainen}}, \bibinfo {author} {\bibfnamefont {G.~C.}\ \bibnamefont
  {McLaughlin}}, \bibinfo {author} {\bibfnamefont {J.~M.}\ \bibnamefont
  {Kelly}}, \bibinfo {author} {\bibfnamefont {S.~N.}\ \bibnamefont {Liddick}},
  \bibinfo {author} {\bibfnamefont {D.~M.}\ \bibnamefont {Lee}}, \bibinfo
  {author} {\bibfnamefont {J.}~\bibnamefont {Lippuner}}, \bibinfo {author}
  {\bibfnamefont {D.}~\bibnamefont {Martin}}, \bibinfo {author} {\bibfnamefont
  {J.}~\bibnamefont {Mendoza-Temis}}, \bibinfo {author} {\bibfnamefont {B.~D.}\
  \bibnamefont {Metzger}}, \bibinfo {author} {\bibfnamefont {M.~R.}\
  \bibnamefont {Mumpower}}, \bibinfo {author} {\bibfnamefont {G.}~\bibnamefont
  {Perdikakis}}, \bibinfo {author} {\bibfnamefont {J.}~\bibnamefont {Pereira}},
  \bibinfo {author} {\bibfnamefont {B.~W.}\ \bibnamefont {O'Shea}}, \bibinfo
  {author} {\bibfnamefont {R.}~\bibnamefont {Reifarth}}, \bibinfo {author}
  {\bibfnamefont {A.~M.}\ \bibnamefont {Rogers}}, \bibinfo {author}
  {\bibfnamefont {D.~M.}\ \bibnamefont {Siegel}}, \bibinfo {author}
  {\bibfnamefont {A.}~\bibnamefont {Spyrou}}, \bibinfo {author} {\bibfnamefont
  {R.}~\bibnamefont {Surman}}, \bibinfo {author} {\bibfnamefont
  {X.}~\bibnamefont {Tang}}, \bibinfo {author} {\bibfnamefont {T.}~\bibnamefont
  {Uesaka}}, \ and\ \bibinfo {author} {\bibfnamefont {M.}~\bibnamefont
  {Wang}},\ }\bibfield  {title} {\enquote {\bibinfo {title} {r-process
  nucleosynthesis: connecting rare-isotope beam facilities with the cosmos},}\
  }\href {\doibase 10.1088/1361-6471/ab0849} {\bibfield  {journal} {\bibinfo
  {journal} {J. Phys. G}\ }\textbf {\bibinfo {volume} {46}},\ \bibinfo {pages}
  {083001} (\bibinfo {year} {2019})}\BibitemShut {NoStop}%
\bibitem [{\citenamefont {Goriely}\ \emph {et~al.}(2013)\citenamefont
  {Goriely}, \citenamefont {Sida}, \citenamefont {Lema{\^{i}}tre},
  \citenamefont {Panebianco}, \citenamefont {Dubray}, \citenamefont {Hilaire},
  \citenamefont {Bauswein},\ and\ \citenamefont {Janka}}]{goriely2013}%
  \BibitemOpen
  \bibfield  {author} {\bibinfo {author} {\bibfnamefont {S.}~\bibnamefont
  {Goriely}}, \bibinfo {author} {\bibfnamefont {J.-L.}\ \bibnamefont {Sida}},
  \bibinfo {author} {\bibfnamefont {J.-F.}\ \bibnamefont {Lema{\^{i}}tre}},
  \bibinfo {author} {\bibfnamefont {S.}~\bibnamefont {Panebianco}}, \bibinfo
  {author} {\bibfnamefont {N.}~\bibnamefont {Dubray}}, \bibinfo {author}
  {\bibfnamefont {S.}~\bibnamefont {Hilaire}}, \bibinfo {author} {\bibfnamefont
  {A.}~\bibnamefont {Bauswein}}, \ and\ \bibinfo {author} {\bibfnamefont
  {H.-T.}\ \bibnamefont {Janka}},\ }\bibfield  {title} {\enquote {\bibinfo
  {title} {New fission fragment distributions and r-process origin of the
  rare-earth elements},}\ }\href {\doibase 10.1103/PhysRevLett.111.242502}
  {\bibfield  {journal} {\bibinfo  {journal} {Phys. Rev. Lett.}\ }\textbf
  {\bibinfo {volume} {111}},\ \bibinfo {pages} {242502} (\bibinfo {year}
  {2013})}\BibitemShut {NoStop}%
\bibitem [{\citenamefont {Eichler}\ \emph {et~al.}(2015)\citenamefont
  {Eichler}, \citenamefont {Arcones}, \citenamefont {Kelic}, \citenamefont
  {Korobkin}, \citenamefont {Langanke}, \citenamefont {Marketin}, \citenamefont
  {Martinez-Pinedo}, \citenamefont {Panov}, \citenamefont {Rauscher},
  \citenamefont {Rosswog}, \citenamefont {Winteler}, \citenamefont {Zinner},\
  and\ \citenamefont {Thielemann}}]{eichler2015}%
  \BibitemOpen
  \bibfield  {author} {\bibinfo {author} {\bibfnamefont {M.}~\bibnamefont
  {Eichler}}, \bibinfo {author} {\bibfnamefont {A.}~\bibnamefont {Arcones}},
  \bibinfo {author} {\bibfnamefont {A.}~\bibnamefont {Kelic}}, \bibinfo
  {author} {\bibfnamefont {O.}~\bibnamefont {Korobkin}}, \bibinfo {author}
  {\bibfnamefont {K.}~\bibnamefont {Langanke}}, \bibinfo {author}
  {\bibfnamefont {T.}~\bibnamefont {Marketin}}, \bibinfo {author}
  {\bibfnamefont {G.}~\bibnamefont {Martinez-Pinedo}}, \bibinfo {author}
  {\bibfnamefont {I.}~\bibnamefont {Panov}}, \bibinfo {author} {\bibfnamefont
  {T.}~\bibnamefont {Rauscher}}, \bibinfo {author} {\bibfnamefont
  {S.}~\bibnamefont {Rosswog}}, \bibinfo {author} {\bibfnamefont
  {C.}~\bibnamefont {Winteler}}, \bibinfo {author} {\bibfnamefont {N.~T.}\
  \bibnamefont {Zinner}}, \ and\ \bibinfo {author} {\bibfnamefont {F.-K.}\
  \bibnamefont {Thielemann}},\ }\bibfield  {title} {\enquote {\bibinfo {title}
  {The role of fission in neutron star mergers and its impact on the r-process
  peaks},}\ }\href {http://stacks.iop.org/0004-637X/808/i=1/a=30} {\bibfield
  {journal} {\bibinfo  {journal} {Astrophys. J.}\ }\textbf {\bibinfo {volume}
  {808}},\ \bibinfo {pages} {30} (\bibinfo {year} {2015})}\BibitemShut
  {NoStop}%
\bibitem [{\citenamefont {Goriely}(2015)}]{goriely2015a}%
  \BibitemOpen
  \bibfield  {author} {\bibinfo {author} {\bibfnamefont {S.}~\bibnamefont
  {Goriely}},\ }\bibfield  {title} {\enquote {\bibinfo {title} {The fundamental
  role of fission during r-process nucleosynthesis in neutron star mergers},}\
  }\href {\doibase 10.1140/epja/i2015-15022-3} {\bibfield  {journal} {\bibinfo
  {journal} {Eur. Phys. J. A}\ }\textbf {\bibinfo {volume} {51}},\ \bibinfo
  {pages} {22} (\bibinfo {year} {2015})}\BibitemShut {NoStop}%
\bibitem [{\citenamefont {Vassh}\ \emph {et~al.}(2019)\citenamefont {Vassh},
  \citenamefont {Vogt}, \citenamefont {Surman}, \citenamefont {Randrup},
  \citenamefont {Sprouse}, \citenamefont {Mumpower}, \citenamefont {Jaffke},
  \citenamefont {Shaw}, \citenamefont {Holmbeck}, \citenamefont {Zhu},\ and\
  \citenamefont {McLaughlin}}]{vassh2019}%
  \BibitemOpen
  \bibfield  {author} {\bibinfo {author} {\bibfnamefont {N.}~\bibnamefont
  {Vassh}}, \bibinfo {author} {\bibfnamefont {R.}~\bibnamefont {Vogt}},
  \bibinfo {author} {\bibfnamefont {R.}~\bibnamefont {Surman}}, \bibinfo
  {author} {\bibfnamefont {J.}~\bibnamefont {Randrup}}, \bibinfo {author}
  {\bibfnamefont {T.~M.}\ \bibnamefont {Sprouse}}, \bibinfo {author}
  {\bibfnamefont {M.~R.}\ \bibnamefont {Mumpower}}, \bibinfo {author}
  {\bibfnamefont {P.}~\bibnamefont {Jaffke}}, \bibinfo {author} {\bibfnamefont
  {D.}~\bibnamefont {Shaw}}, \bibinfo {author} {\bibfnamefont {E.~M.}\
  \bibnamefont {Holmbeck}}, \bibinfo {author} {\bibfnamefont {Y.-L.}\
  \bibnamefont {Zhu}}, \ and\ \bibinfo {author} {\bibfnamefont {G.~C.}\
  \bibnamefont {McLaughlin}},\ }\bibfield  {title} {\enquote {\bibinfo {title}
  {Using excitation-energy dependent fission yields to identify key fissioning
  nuclei in r -process nucleosynthesis},}\ }\href {\doibase
  10.1088/1361-6471/ab0bea} {\bibfield  {journal} {\bibinfo  {journal} {J.
  Phys. G}\ }\textbf {\bibinfo {volume} {46}},\ \bibinfo {pages} {065202}
  (\bibinfo {year} {2019})}\BibitemShut {NoStop}%
\bibitem [{\citenamefont {Vassh}\ \emph {et~al.}(2020)\citenamefont {Vassh},
  \citenamefont {Mumpower}, \citenamefont {McLaughlin}, \citenamefont
  {Sprouse},\ and\ \citenamefont {Surman}}]{vassh2019a}%
  \BibitemOpen
  \bibfield  {author} {\bibinfo {author} {\bibfnamefont {N.}~\bibnamefont
  {Vassh}}, \bibinfo {author} {\bibfnamefont {M.~R.}\ \bibnamefont {Mumpower}},
  \bibinfo {author} {\bibfnamefont {G.~C.}\ \bibnamefont {McLaughlin}},
  \bibinfo {author} {\bibfnamefont {T.~M.}\ \bibnamefont {Sprouse}}, \ and\
  \bibinfo {author} {\bibfnamefont {R.}~\bibnamefont {Surman}},\ }\bibfield
  {title} {\enquote {\bibinfo {title} {{Coproduction of Light and Heavy r
  -process Elements via Fission Deposition}},}\ }\href {\doibase
  10.3847/1538-4357/ab91a9} {\bibfield  {journal} {\bibinfo  {journal} {The
  Astrophysical Journal}\ }\textbf {\bibinfo {volume} {896}},\ \bibinfo {pages}
  {28} (\bibinfo {year} {2020})},\ \Eprint {http://arxiv.org/abs/1911.07766}
  {1911.07766} \BibitemShut {NoStop}%
\bibitem [{\citenamefont {C{\^{o}}t{\'{e}}}\ \emph {et~al.}(2018)\citenamefont
  {C{\^{o}}t{\'{e}}}, \citenamefont {Fryer}, \citenamefont {Belczynski},
  \citenamefont {Korobkin}, \citenamefont {Chru{\'{s}}li{\'{n}}ska},
  \citenamefont {Vassh}, \citenamefont {Mumpower}, \citenamefont {Lippuner},
  \citenamefont {Sprouse}, \citenamefont {Surman},\ and\ \citenamefont
  {Wollaeger}}]{cote2017}%
  \BibitemOpen
  \bibfield  {author} {\bibinfo {author} {\bibfnamefont {B.}~\bibnamefont
  {C{\^{o}}t{\'{e}}}}, \bibinfo {author} {\bibfnamefont {C.~L.}\ \bibnamefont
  {Fryer}}, \bibinfo {author} {\bibfnamefont {K.}~\bibnamefont {Belczynski}},
  \bibinfo {author} {\bibfnamefont {O.}~\bibnamefont {Korobkin}}, \bibinfo
  {author} {\bibfnamefont {M.}~\bibnamefont {Chru{\'{s}}li{\'{n}}ska}},
  \bibinfo {author} {\bibfnamefont {N.}~\bibnamefont {Vassh}}, \bibinfo
  {author} {\bibfnamefont {M.~R.}\ \bibnamefont {Mumpower}}, \bibinfo {author}
  {\bibfnamefont {J.}~\bibnamefont {Lippuner}}, \bibinfo {author}
  {\bibfnamefont {T.~M.}\ \bibnamefont {Sprouse}}, \bibinfo {author}
  {\bibfnamefont {R.}~\bibnamefont {Surman}}, \ and\ \bibinfo {author}
  {\bibfnamefont {R.}~\bibnamefont {Wollaeger}},\ }\bibfield  {title} {\enquote
  {\bibinfo {title} {{The Origin of r -process Elements in the Milky Way}},}\
  }\href {\doibase 10.3847/1538-4357/aaad67} {\bibfield  {journal} {\bibinfo
  {journal} {The Astrophysical Journal}\ }\textbf {\bibinfo {volume} {855}},\
  \bibinfo {pages} {99} (\bibinfo {year} {2018})},\ \Eprint
  {http://arxiv.org/abs/1710.05875} {1710.05875} \BibitemShut {NoStop}%
\bibitem [{\citenamefont {Sadhukhan}\ \emph {et~al.}(2016)\citenamefont
  {Sadhukhan}, \citenamefont {Nazarewicz},\ and\ \citenamefont
  {Schunck}}]{sadhukhan2016}%
  \BibitemOpen
  \bibfield  {author} {\bibinfo {author} {\bibfnamefont {J.}~\bibnamefont
  {Sadhukhan}}, \bibinfo {author} {\bibfnamefont {W.}~\bibnamefont
  {Nazarewicz}}, \ and\ \bibinfo {author} {\bibfnamefont {N.}~\bibnamefont
  {Schunck}},\ }\bibfield  {title} {\enquote {\bibinfo {title} {Microscopic
  modeling of mass and charge distributions in the spontaneous fission of
  $^{240}${Pu}},}\ }\href {\doibase 10.1103/PhysRevC.93.011304} {\bibfield
  {journal} {\bibinfo  {journal} {Phys. Rev. C}\ }\textbf {\bibinfo {volume}
  {93}},\ \bibinfo {pages} {011304} (\bibinfo {year} {2016})}\BibitemShut
  {NoStop}%
\bibitem [{\citenamefont {Sierk}(2017)}]{sierk2017}%
  \BibitemOpen
  \bibfield  {author} {\bibinfo {author} {\bibfnamefont {A.~J.}\ \bibnamefont
  {Sierk}},\ }\bibfield  {title} {\enquote {\bibinfo {title} {Langevin model of
  low-energy fission},}\ }\href {\doibase 10.1103/PhysRevC.96.034603}
  {\bibfield  {journal} {\bibinfo  {journal} {Phys. Rev. C}\ }\textbf {\bibinfo
  {volume} {96}},\ \bibinfo {pages} {034603} (\bibinfo {year}
  {2017})}\BibitemShut {NoStop}%
\bibitem [{\citenamefont {Usang}\ \emph {et~al.}(2019)\citenamefont {Usang},
  \citenamefont {Ivanyuk}, \citenamefont {Ishizuka},\ and\ \citenamefont
  {Chiba}}]{usang2019}%
  \BibitemOpen
  \bibfield  {author} {\bibinfo {author} {\bibfnamefont {M.~D.}\ \bibnamefont
  {Usang}}, \bibinfo {author} {\bibfnamefont {F.~A.}\ \bibnamefont {Ivanyuk}},
  \bibinfo {author} {\bibfnamefont {C.}~\bibnamefont {Ishizuka}}, \ and\
  \bibinfo {author} {\bibfnamefont {S.}~\bibnamefont {Chiba}},\ }\bibfield
  {title} {\enquote {\bibinfo {title} {{Correlated transitions in TKE and mass
  distributions of fission fragments described by 4-D Langevin equation}},}\
  }\href {\doibase 10.1038/s41598-018-37993-7} {\bibfield  {journal} {\bibinfo
  {journal} {Sci. Rep.}\ }\textbf {\bibinfo {volume} {9}},\ \bibinfo {pages}
  {1525} (\bibinfo {year} {2019})}\BibitemShut {NoStop}%
\bibitem [{\citenamefont {Simenel}\ and\ \citenamefont
  {Umar}(2018)}]{simenel2018}%
  \BibitemOpen
  \bibfield  {author} {\bibinfo {author} {\bibfnamefont {C.}~\bibnamefont
  {Simenel}}\ and\ \bibinfo {author} {\bibfnamefont {A.~S.}\ \bibnamefont
  {Umar}},\ }\bibfield  {title} {\enquote {\bibinfo {title} {Heavy-ion
  collisions and fission dynamics with the time-dependent {Hartree-Fock} theory
  and its extensions},}\ }\href {\doibase 10.1016/j.ppnp.2018.07.002}
  {\bibfield  {journal} {\bibinfo  {journal} {Prog. Part. Nucl. Phys.}\
  }\textbf {\bibinfo {volume} {103}},\ \bibinfo {pages} {19--66} (\bibinfo
  {year} {2018})}\BibitemShut {NoStop}%
\bibitem [{\citenamefont {Scamps}\ \emph {et~al.}(2015)\citenamefont {Scamps},
  \citenamefont {Simenel},\ and\ \citenamefont {Lacroix}}]{scamps2015a}%
  \BibitemOpen
  \bibfield  {author} {\bibinfo {author} {\bibfnamefont {G.}~\bibnamefont
  {Scamps}}, \bibinfo {author} {\bibfnamefont {C.}~\bibnamefont {Simenel}}, \
  and\ \bibinfo {author} {\bibfnamefont {D.}~\bibnamefont {Lacroix}},\
  }\bibfield  {title} {\enquote {\bibinfo {title} {Dynamical description of the
  fission process using the {TD-BCS} theory},}\ }\href {\doibase
  10.1063/1.4932264} {\bibfield  {journal} {\bibinfo  {journal} {AIP Conf.
  Proc.}\ }\textbf {\bibinfo {volume} {1681}},\ \bibinfo {pages} {040003}
  (\bibinfo {year} {2015})}\BibitemShut {NoStop}%
\bibitem [{\citenamefont {Bulgac}\ \emph
  {et~al.}(2019{\natexlab{a}})\citenamefont {Bulgac}, \citenamefont {Jin},
  \citenamefont {Roche}, \citenamefont {Schunck},\ and\ \citenamefont
  {Stetcu}}]{bulgac2018}%
  \BibitemOpen
  \bibfield  {author} {\bibinfo {author} {\bibfnamefont {A.}~\bibnamefont
  {Bulgac}}, \bibinfo {author} {\bibfnamefont {S.}~\bibnamefont {Jin}},
  \bibinfo {author} {\bibfnamefont {K.~J.}\ \bibnamefont {Roche}}, \bibinfo
  {author} {\bibfnamefont {N.}~\bibnamefont {Schunck}}, \ and\ \bibinfo
  {author} {\bibfnamefont {I.}~\bibnamefont {Stetcu}},\ }\bibfield  {title}
  {\enquote {\bibinfo {title} {Fission dynamics of $^{240}\mathrm{Pu}$ from
  saddle to scission and beyond},}\ }\href {\doibase
  10.1103/PhysRevC.100.034615} {\bibfield  {journal} {\bibinfo  {journal}
  {Phys. Rev. C}\ }\textbf {\bibinfo {volume} {100}},\ \bibinfo {pages}
  {034615} (\bibinfo {year} {2019}{\natexlab{a}})}\BibitemShut {NoStop}%
\bibitem [{\citenamefont {Bulgac}\ \emph
  {et~al.}(2019{\natexlab{b}})\citenamefont {Bulgac}, \citenamefont {Jin},\
  and\ \citenamefont {Stetcu}}]{bulgac2019}%
  \BibitemOpen
  \bibfield  {author} {\bibinfo {author} {\bibfnamefont {A.}~\bibnamefont
  {Bulgac}}, \bibinfo {author} {\bibfnamefont {S.}~\bibnamefont {Jin}}, \ and\
  \bibinfo {author} {\bibfnamefont {I.}~\bibnamefont {Stetcu}},\ }\bibfield
  {title} {\enquote {\bibinfo {title} {Unitary evolution with fluctuations and
  dissipation},}\ }\href {\doibase 10.1103/PhysRevC.100.014615} {\bibfield
  {journal} {\bibinfo  {journal} {Phys. Rev. C}\ }\textbf {\bibinfo {volume}
  {100}},\ \bibinfo {pages} {014615} (\bibinfo {year}
  {2019}{\natexlab{b}})}\BibitemShut {NoStop}%
\bibitem [{\citenamefont {Regnier}\ \emph {et~al.}(2019)\citenamefont
  {Regnier}, \citenamefont {Dubray},\ and\ \citenamefont
  {Schunck}}]{regnier2019}%
  \BibitemOpen
  \bibfield  {author} {\bibinfo {author} {\bibfnamefont {D.}~\bibnamefont
  {Regnier}}, \bibinfo {author} {\bibfnamefont {N.}~\bibnamefont {Dubray}}, \
  and\ \bibinfo {author} {\bibfnamefont {N.}~\bibnamefont {Schunck}},\
  }\bibfield  {title} {\enquote {\bibinfo {title} {{From asymmetric to
  symmetric fission in the fermium isotopes within the time-dependent
  generator-coordinate-method formalism}},}\ }\href {\doibase
  10.1103/PhysRevC.99.024611} {\bibfield  {journal} {\bibinfo  {journal} {Phys.
  Rev. C}\ }\textbf {\bibinfo {volume} {99}},\ \bibinfo {pages} {024611}
  (\bibinfo {year} {2019})}\BibitemShut {NoStop}%
\bibitem [{\citenamefont {Zhao}\ \emph {et~al.}(2019)\citenamefont {Zhao},
  \citenamefont {Xiang}, \citenamefont {Li}, \citenamefont {Nik{\v s}i{\'c}},
  \citenamefont {Vretenar},\ and\ \citenamefont {Zhou}}]{zhao2019}%
  \BibitemOpen
  \bibfield  {author} {\bibinfo {author} {\bibfnamefont {J.}~\bibnamefont
  {Zhao}}, \bibinfo {author} {\bibfnamefont {J.}~\bibnamefont {Xiang}},
  \bibinfo {author} {\bibfnamefont {Z.-P.}\ \bibnamefont {Li}}, \bibinfo
  {author} {\bibfnamefont {T.}~\bibnamefont {Nik{\v s}i{\'c}}}, \bibinfo
  {author} {\bibfnamefont {D.}~\bibnamefont {Vretenar}}, \ and\ \bibinfo
  {author} {\bibfnamefont {S.-G.}\ \bibnamefont {Zhou}},\ }\bibfield  {title}
  {\enquote {\bibinfo {title} {Time-dependent generator-coordinate-method study
  of mass-asymmetric fission of actinides},}\ }\href {\doibase
  10.1103/PhysRevC.99.054613} {\bibfield  {journal} {\bibinfo  {journal} {Phys.
  Rev. C}\ }\textbf {\bibinfo {volume} {99}},\ \bibinfo {pages} {054613}
  (\bibinfo {year} {2019})}\BibitemShut {NoStop}%
\bibitem [{\citenamefont {Schmidt}\ and\ \citenamefont
  {Jurado}(2018)}]{schmidt2018}%
  \BibitemOpen
  \bibfield  {author} {\bibinfo {author} {\bibfnamefont {K.-H.}\ \bibnamefont
  {Schmidt}}\ and\ \bibinfo {author} {\bibfnamefont {B.}~\bibnamefont
  {Jurado}},\ }\bibfield  {title} {\enquote {\bibinfo {title} {Review on the
  progress in nuclear fission—experimental methods and theoretical
  descriptions},}\ }\href {\doibase 10.1088/1361-6633/aacfa7} {\bibfield
  {journal} {\bibinfo  {journal} {Rep. Prog. Phys.}\ }\textbf {\bibinfo
  {volume} {81}},\ \bibinfo {pages} {106301} (\bibinfo {year}
  {2018})}\BibitemShut {NoStop}%
\bibitem [{\citenamefont {Randrup}\ \emph {et~al.}(2011)\citenamefont
  {Randrup}, \citenamefont {M{\"{o}}ller},\ and\ \citenamefont
  {Sierk}}]{randrup2011}%
  \BibitemOpen
  \bibfield  {author} {\bibinfo {author} {\bibfnamefont {J.}~\bibnamefont
  {Randrup}}, \bibinfo {author} {\bibfnamefont {P.}~\bibnamefont
  {M{\"{o}}ller}}, \ and\ \bibinfo {author} {\bibfnamefont {A.~J.}\
  \bibnamefont {Sierk}},\ }\bibfield  {title} {\enquote {\bibinfo {title}
  {Fission-fragment mass distributions from strongly damped shape evolution},}\
  }\href {\doibase 10.1103/PhysRevC.84.034613} {\bibfield  {journal} {\bibinfo
  {journal} {Phys. Rev. C}\ }\textbf {\bibinfo {volume} {84}},\ \bibinfo
  {pages} {034613} (\bibinfo {year} {2011})}\BibitemShut {NoStop}%
\bibitem [{\citenamefont {M{\"o}ller}\ and\ \citenamefont
  {Ichikawa}(2015)}]{moller2015c}%
  \BibitemOpen
  \bibfield  {author} {\bibinfo {author} {\bibfnamefont {P.}~\bibnamefont
  {M{\"o}ller}}\ and\ \bibinfo {author} {\bibfnamefont {T.}~\bibnamefont
  {Ichikawa}},\ }\bibfield  {title} {\enquote {\bibinfo {title} {A method to
  calculate fission-fragment yields {Y(Z,N)} versus proton and neutron number
  in the brownian shape-motion model},}\ }\href {\doibase
  10.1140/epja/i2015-15173-1} {\bibfield  {journal} {\bibinfo  {journal} {Eur.
  Phys. J. A}\ }\textbf {\bibinfo {volume} {51}},\ \bibinfo {pages} {173}
  (\bibinfo {year} {2015})}\BibitemShut {NoStop}%
\bibitem [{\citenamefont {Mumpower}\ \emph {et~al.}(2019)\citenamefont
  {Mumpower}, \citenamefont {Jaffke}, \citenamefont {Verriere},\ and\
  \citenamefont {Randrup}}]{mumpower2019}%
  \BibitemOpen
  \bibfield  {author} {\bibinfo {author} {\bibfnamefont {M.~R.}\ \bibnamefont
  {Mumpower}}, \bibinfo {author} {\bibfnamefont {P.}~\bibnamefont {Jaffke}},
  \bibinfo {author} {\bibfnamefont {M.}~\bibnamefont {Verriere}}, \ and\
  \bibinfo {author} {\bibfnamefont {J.}~\bibnamefont {Randrup}},\ }\bibfield
  {title} {\enquote {\bibinfo {title} {Primary fission fragment mass yields
  across the chart of nuclides},}\ }\href {http://arxiv.org/abs/1911.06344} {\
  (\bibinfo {year} {2019})},\ \Eprint {http://arxiv.org/abs/1911.06344}
  {arXiv:1911.06344 [nucl-th]} \BibitemShut {NoStop}%
\bibitem [{\citenamefont {Fong}(1953)}]{fong1953}%
  \BibitemOpen
  \bibfield  {author} {\bibinfo {author} {\bibfnamefont {P.}~\bibnamefont
  {Fong}},\ }\bibfield  {title} {\enquote {\bibinfo {title} {Asymmetric
  fission},}\ }\href {\doibase 10.1103/PhysRev.89.332} {\bibfield  {journal}
  {\bibinfo  {journal} {Phys. Rev.}\ }\textbf {\bibinfo {volume} {89}},\
  \bibinfo {pages} {332--333} (\bibinfo {year} {1953})}\BibitemShut {NoStop}%
\bibitem [{\citenamefont {Erba}\ \emph {et~al.}(1966)\citenamefont {Erba},
  \citenamefont {Facchini},\ and\ \citenamefont
  {Saetta-Menichella}}]{Erba1966}%
  \BibitemOpen
  \bibfield  {author} {\bibinfo {author} {\bibfnamefont {E.}~\bibnamefont
  {Erba}}, \bibinfo {author} {\bibfnamefont {U.}~\bibnamefont {Facchini}}, \
  and\ \bibinfo {author} {\bibfnamefont {E.}~\bibnamefont
  {Saetta-Menichella}},\ }\bibfield  {title} {\enquote {\bibinfo {title}
  {Statistical analysis of low-energy fission},}\ }\href {\doibase
  10.1016/0029-5582(66)91017-0} {\bibfield  {journal} {\bibinfo  {journal}
  {Nucl. Phys.}\ }\textbf {\bibinfo {volume} {84}},\ \bibinfo {pages} {595 --
  608} (\bibinfo {year} {1966})}\BibitemShut {NoStop}%
\bibitem [{\citenamefont {Wilkins}\ \emph {et~al.}(1976)\citenamefont
  {Wilkins}, \citenamefont {Steinberg},\ and\ \citenamefont
  {Chasman}}]{wilkins1976}%
  \BibitemOpen
  \bibfield  {author} {\bibinfo {author} {\bibfnamefont {B.~D.}\ \bibnamefont
  {Wilkins}}, \bibinfo {author} {\bibfnamefont {E.~P.}\ \bibnamefont
  {Steinberg}}, \ and\ \bibinfo {author} {\bibfnamefont {R.~R.}\ \bibnamefont
  {Chasman}},\ }\bibfield  {title} {\enquote {\bibinfo {title} {Scission-point
  model of nuclear fission based on deformed-shell effects},}\ }\href {\doibase
  10.1103/PhysRevC.14.1832} {\bibfield  {journal} {\bibinfo  {journal} {Phys.
  Rev. C}\ }\textbf {\bibinfo {volume} {14}},\ \bibinfo {pages} {1832--1863}
  (\bibinfo {year} {1976})}\BibitemShut {NoStop}%
\bibitem [{\citenamefont {Lema{\^{i}}tre}\ \emph {et~al.}(2015)\citenamefont
  {Lema{\^{i}}tre}, \citenamefont {Panebianco}, \citenamefont {Sida},
  \citenamefont {Hilaire},\ and\ \citenamefont {Heinrich}}]{lemaitre2015}%
  \BibitemOpen
  \bibfield  {author} {\bibinfo {author} {\bibfnamefont {J.-F.}\ \bibnamefont
  {Lema{\^{i}}tre}}, \bibinfo {author} {\bibfnamefont {S.}~\bibnamefont
  {Panebianco}}, \bibinfo {author} {\bibfnamefont {J.-L.}\ \bibnamefont
  {Sida}}, \bibinfo {author} {\bibfnamefont {S.}~\bibnamefont {Hilaire}}, \
  and\ \bibinfo {author} {\bibfnamefont {S.}~\bibnamefont {Heinrich}},\
  }\bibfield  {title} {\enquote {\bibinfo {title} {{New statistical
  scission-point model to predict fission fragment observables}},}\ }\href
  {\doibase 10.1103/PhysRevC.92.034617} {\bibfield  {journal} {\bibinfo
  {journal} {Phys. Rev. C}\ }\textbf {\bibinfo {volume} {92}},\ \bibinfo
  {pages} {034617} (\bibinfo {year} {2015})}\BibitemShut {NoStop}%
\bibitem [{\citenamefont {Lema{\^{i}}tre}\ \emph {et~al.}(2019)\citenamefont
  {Lema{\^{i}}tre}, \citenamefont {Goriely}, \citenamefont {Hilaire},\ and\
  \citenamefont {Sida}}]{lemaitre2019}%
  \BibitemOpen
  \bibfield  {author} {\bibinfo {author} {\bibfnamefont {J.-F.}\ \bibnamefont
  {Lema{\^{i}}tre}}, \bibinfo {author} {\bibfnamefont {S.}~\bibnamefont
  {Goriely}}, \bibinfo {author} {\bibfnamefont {S.}~\bibnamefont {Hilaire}}, \
  and\ \bibinfo {author} {\bibfnamefont {J.-L.}\ \bibnamefont {Sida}},\
  }\bibfield  {title} {\enquote {\bibinfo {title} {{Fully microscopic
  scission-point model to predict fission fragment observables}},}\ }\href
  {\doibase 10.1103/PhysRevC.99.034612} {\bibfield  {journal} {\bibinfo
  {journal} {Phys. Rev. C}\ }\textbf {\bibinfo {volume} {99}},\ \bibinfo
  {pages} {034612} (\bibinfo {year} {2019})}\BibitemShut {NoStop}%
\bibitem [{\citenamefont {Andreev}\ \emph {et~al.}(2006)\citenamefont
  {Andreev}, \citenamefont {Adamian}, \citenamefont {Antonenko}, \citenamefont
  {Ivanova}, \citenamefont {Kuklin},\ and\ \citenamefont
  {Scheid}}]{Andreev2006}%
  \BibitemOpen
  \bibfield  {author} {\bibinfo {author} {\bibfnamefont {A.~V.}\ \bibnamefont
  {Andreev}}, \bibinfo {author} {\bibfnamefont {G.~G.}\ \bibnamefont
  {Adamian}}, \bibinfo {author} {\bibfnamefont {N.~V.}\ \bibnamefont
  {Antonenko}}, \bibinfo {author} {\bibfnamefont {S.~P.}\ \bibnamefont
  {Ivanova}}, \bibinfo {author} {\bibfnamefont {S.~N.}\ \bibnamefont {Kuklin}},
  \ and\ \bibinfo {author} {\bibfnamefont {W.}~\bibnamefont {Scheid}},\
  }\bibfield  {title} {\enquote {\bibinfo {title} {Ternary fission within
  statistical approach},}\ }\href {\doibase 10.1140/epja/i2006-10145-2}
  {\bibfield  {journal} {\bibinfo  {journal} {Eur. Phys. J. A}\ }\textbf
  {\bibinfo {volume} {30}},\ \bibinfo {pages} {579--589} (\bibinfo {year}
  {2006})}\BibitemShut {NoStop}%
\bibitem [{\citenamefont {Pa\ifmmode~\mbox{\c{s}}\else \c{s}\fi{}ca}\ \emph
  {et~al.}(2019)\citenamefont {Pa\ifmmode~\mbox{\c{s}}\else \c{s}\fi{}ca},
  \citenamefont {Andreev}, \citenamefont {Adamian},\ and\ \citenamefont
  {Antonenko}}]{Pasca2019}%
  \BibitemOpen
  \bibfield  {author} {\bibinfo {author} {\bibfnamefont {H.}~\bibnamefont
  {Pa\ifmmode~\mbox{\c{s}}\else \c{s}\fi{}ca}}, \bibinfo {author}
  {\bibfnamefont {A.~V.}\ \bibnamefont {Andreev}}, \bibinfo {author}
  {\bibfnamefont {G.~G.}\ \bibnamefont {Adamian}}, \ and\ \bibinfo {author}
  {\bibfnamefont {N.~V.}\ \bibnamefont {Antonenko}},\ }\bibfield  {title}
  {\enquote {\bibinfo {title} {Change of the shape of mass and charge
  distributions in fission of cf isotopes with excitation energy},}\ }\href
  {\doibase 10.1103/PhysRevC.99.064611} {\bibfield  {journal} {\bibinfo
  {journal} {Phys. Rev. C}\ }\textbf {\bibinfo {volume} {99}},\ \bibinfo
  {pages} {064611} (\bibinfo {year} {2019})}\BibitemShut {NoStop}%
\bibitem [{\citenamefont {Brosa}\ and\ \citenamefont
  {Grossmann}(1983)}]{brosa1983}%
  \BibitemOpen
  \bibfield  {author} {\bibinfo {author} {\bibfnamefont {U.}~\bibnamefont
  {Brosa}}\ and\ \bibinfo {author} {\bibfnamefont {S.}~\bibnamefont
  {Grossmann}},\ }\bibfield  {title} {\enquote {\bibinfo {title} {In the exit
  channel of nuclear fission},}\ }\href {\doibase 10.1007/BF01415223}
  {\bibfield  {journal} {\bibinfo  {journal} {Z. Phys. A}\ }\textbf {\bibinfo
  {volume} {310}},\ \bibinfo {pages} {177--187} (\bibinfo {year}
  {1983})}\BibitemShut {NoStop}%
\bibitem [{\citenamefont {Oberstedt}\ \emph {et~al.}(1998)\citenamefont
  {Oberstedt}, \citenamefont {Hambsch},\ and\ \citenamefont
  {Viv{\`e}s}}]{Ober1998}%
  \BibitemOpen
  \bibfield  {author} {\bibinfo {author} {\bibfnamefont {S.}~\bibnamefont
  {Oberstedt}}, \bibinfo {author} {\bibfnamefont {F.-J.}\ \bibnamefont
  {Hambsch}}, \ and\ \bibinfo {author} {\bibfnamefont {F.}~\bibnamefont
  {Viv{\`e}s}},\ }\bibfield  {title} {\enquote {\bibinfo {title} {Fission-mode
  calculations for $^{239}${U}, a revision of the multi-modal random
  neck-rupture model},}\ }\href {\doibase 10.1016/S0375-9474(98)00598-3}
  {\bibfield  {journal} {\bibinfo  {journal} {Nucl. Phys. A}\ }\textbf
  {\bibinfo {volume} {644}},\ \bibinfo {pages} {289 -- 305} (\bibinfo {year}
  {1998})}\BibitemShut {NoStop}%
\bibitem [{\citenamefont {Schmidt}\ \emph {et~al.}(2016)\citenamefont
  {Schmidt}, \citenamefont {Jurado}, \citenamefont {Amouroux},\ and\
  \citenamefont {Schmitt}}]{schmidt2016}%
  \BibitemOpen
  \bibfield  {author} {\bibinfo {author} {\bibfnamefont {K.-H.}\ \bibnamefont
  {Schmidt}}, \bibinfo {author} {\bibfnamefont {B.}~\bibnamefont {Jurado}},
  \bibinfo {author} {\bibfnamefont {C.}~\bibnamefont {Amouroux}}, \ and\
  \bibinfo {author} {\bibfnamefont {C.}~\bibnamefont {Schmitt}},\ }\bibfield
  {title} {\enquote {\bibinfo {title} {General description of fission
  observables: {GEF} model code},}\ }\href {\doibase 10.1016/j.nds.2015.12.009}
  {\bibfield  {journal} {\bibinfo  {journal} {Nucl. Data Sheets}\ }\textbf
  {\bibinfo {volume} {131}},\ \bibinfo {pages} {107--221} (\bibinfo {year}
  {2016})}\BibitemShut {NoStop}%
\bibitem [{\citenamefont {Keli{\'c}}\ \emph {et~al.}(2009)\citenamefont
  {Keli{\'c}}, \citenamefont {Ricciardi},\ and\ \citenamefont
  {Schmidt}}]{kelic2009}%
  \BibitemOpen
  \bibfield  {author} {\bibinfo {author} {\bibfnamefont {A.}~\bibnamefont
  {Keli{\'c}}}, \bibinfo {author} {\bibfnamefont {M.~V.}\ \bibnamefont
  {Ricciardi}}, \ and\ \bibinfo {author} {\bibfnamefont {K.-H.}\ \bibnamefont
  {Schmidt}},\ }\bibfield  {title} {\enquote {\bibinfo {title} {{ABLA07} -
  towards a complete description of the decay channels of a nuclear system from
  spontaneous fission to multifragmentation},}\ }\href@noop {} {\  (\bibinfo
  {year} {2009})},\ \Eprint {http://arxiv.org/abs/0906.4193} {arXiv:0906.4193
  [nucl-th]} \BibitemShut {NoStop}%
\bibitem [{\citenamefont {Mendoza-Temis}\ \emph {et~al.}(2015)\citenamefont
  {Mendoza-Temis}, \citenamefont {Wu}, \citenamefont {Langanke}, \citenamefont
  {Mart{\'{i}}nez-Pinedo}, \citenamefont {Bauswein},\ and\ \citenamefont
  {Janka}}]{mendoza2015}%
  \BibitemOpen
  \bibfield  {author} {\bibinfo {author} {\bibfnamefont {J.~d.~J.}\
  \bibnamefont {Mendoza-Temis}}, \bibinfo {author} {\bibfnamefont {M.-R.}\
  \bibnamefont {Wu}}, \bibinfo {author} {\bibfnamefont {K.}~\bibnamefont
  {Langanke}}, \bibinfo {author} {\bibfnamefont {G.}~\bibnamefont
  {Mart{\'{i}}nez-Pinedo}}, \bibinfo {author} {\bibfnamefont {A.}~\bibnamefont
  {Bauswein}}, \ and\ \bibinfo {author} {\bibfnamefont {H.-T.}\ \bibnamefont
  {Janka}},\ }\bibfield  {title} {\enquote {\bibinfo {title} {Nuclear
  robustness of the r process in neutron-star mergers},}\ }\href {\doibase
  10.1103/PhysRevC.92.055805} {\bibfield  {journal} {\bibinfo  {journal} {Phys.
  Rev. C}\ }\textbf {\bibinfo {volume} {92}},\ \bibinfo {pages} {055805}
  (\bibinfo {year} {2015})}\BibitemShut {NoStop}%
\bibitem [{\citenamefont {Goriely}\ and\ \citenamefont {{Mart{\'{i}}nez
  Pinedo}}(2015)}]{goriely2015}%
  \BibitemOpen
  \bibfield  {author} {\bibinfo {author} {\bibfnamefont {S.}~\bibnamefont
  {Goriely}}\ and\ \bibinfo {author} {\bibfnamefont {G.}~\bibnamefont
  {{Mart{\'{i}}nez Pinedo}}},\ }\bibfield  {title} {\enquote {\bibinfo {title}
  {The production of transuranium elements by the r-process nucleosynthesis},}\
  }\href {\doibase 10.1016/j.nuclphysa.2015.07.020} {\bibfield  {journal}
  {\bibinfo  {journal} {Nucl. Phys. A}\ }\textbf {\bibinfo {volume} {944}},\
  \bibinfo {pages} {158--176} (\bibinfo {year} {2015})}\BibitemShut {NoStop}%
\bibitem [{\citenamefont {Sadhukhan}\ \emph {et~al.}(2017)\citenamefont
  {Sadhukhan}, \citenamefont {Zhang}, \citenamefont {Nazarewicz},\ and\
  \citenamefont {Schunck}}]{sadhukhan2017}%
  \BibitemOpen
  \bibfield  {author} {\bibinfo {author} {\bibfnamefont {J.}~\bibnamefont
  {Sadhukhan}}, \bibinfo {author} {\bibfnamefont {C.}~\bibnamefont {Zhang}},
  \bibinfo {author} {\bibfnamefont {W.}~\bibnamefont {Nazarewicz}}, \ and\
  \bibinfo {author} {\bibfnamefont {N.}~\bibnamefont {Schunck}},\ }\bibfield
  {title} {\enquote {\bibinfo {title} {Formation and distribution of fragments
  in the spontaneous fission of $^{240}${Pu}},}\ }\href {\doibase
  10.1103/PhysRevC.96.061301} {\bibfield  {journal} {\bibinfo  {journal} {Phys.
  Rev. C}\ }\textbf {\bibinfo {volume} {96}},\ \bibinfo {pages} {061301}
  (\bibinfo {year} {2017})}\BibitemShut {NoStop}%
\bibitem [{\citenamefont {Matheson}\ \emph {et~al.}(2019)\citenamefont
  {Matheson}, \citenamefont {Giuliani}, \citenamefont {Nazarewicz},
  \citenamefont {Sadhukhan},\ and\ \citenamefont {Schunck}}]{matheson2019}%
  \BibitemOpen
  \bibfield  {author} {\bibinfo {author} {\bibfnamefont {Z.}~\bibnamefont
  {Matheson}}, \bibinfo {author} {\bibfnamefont {S.~A.}\ \bibnamefont
  {Giuliani}}, \bibinfo {author} {\bibfnamefont {W.}~\bibnamefont
  {Nazarewicz}}, \bibinfo {author} {\bibfnamefont {J.}~\bibnamefont
  {Sadhukhan}}, \ and\ \bibinfo {author} {\bibfnamefont {N.}~\bibnamefont
  {Schunck}},\ }\bibfield  {title} {\enquote {\bibinfo {title} {Cluster
  radioactivity of $^{294}${Og}},}\ }\href {\doibase
  10.1103/PhysRevC.99.041304} {\bibfield  {journal} {\bibinfo  {journal} {Phys.
  Rev. C}\ }\textbf {\bibinfo {volume} {99}},\ \bibinfo {pages} {041304}
  (\bibinfo {year} {2019})}\BibitemShut {NoStop}%
\bibitem [{\citenamefont {Negele}(1989)}]{negele1989}%
  \BibitemOpen
  \bibfield  {author} {\bibinfo {author} {\bibfnamefont {J.~W.}\ \bibnamefont
  {Negele}},\ }\bibfield  {title} {\enquote {\bibinfo {title} {Microscopic
  theory of fission dynamics},}\ }\href {\doibase
  https://doi.org/10.1016/0375-9474(89)90676-3} {\bibfield  {journal} {\bibinfo
   {journal} {Nucl. Phys. A}\ }\textbf {\bibinfo {volume} {502}},\ \bibinfo
  {pages} {371 -- 386} (\bibinfo {year} {1989})}\BibitemShut {NoStop}%
\bibitem [{\citenamefont {Nazarewicz}(1993)}]{nazarewicz1993}%
  \BibitemOpen
  \bibfield  {author} {\bibinfo {author} {\bibfnamefont {W.}~\bibnamefont
  {Nazarewicz}},\ }\bibfield  {title} {\enquote {\bibinfo {title} {Diabaticity
  of nuclear motion: problems and perspectives},}\ }\href {\doibase
  10.1016/0375-9474(93)90565-F} {\bibfield  {journal} {\bibinfo  {journal}
  {Nucl. Phys. A}\ }\textbf {\bibinfo {volume} {557}},\ \bibinfo {pages}
  {489--514} (\bibinfo {year} {1993})}\BibitemShut {NoStop}%
\bibitem [{\citenamefont {Sadhukhan}\ \emph {et~al.}(2013)\citenamefont
  {Sadhukhan}, \citenamefont {Mazurek}, \citenamefont {Baran}, \citenamefont
  {Dobaczewski}, \citenamefont {Nazarewicz},\ and\ \citenamefont
  {Sheikh}}]{sadhukhan2013}%
  \BibitemOpen
  \bibfield  {author} {\bibinfo {author} {\bibfnamefont {J.}~\bibnamefont
  {Sadhukhan}}, \bibinfo {author} {\bibfnamefont {K.}~\bibnamefont {Mazurek}},
  \bibinfo {author} {\bibfnamefont {A.}~\bibnamefont {Baran}}, \bibinfo
  {author} {\bibfnamefont {J.}~\bibnamefont {Dobaczewski}}, \bibinfo {author}
  {\bibfnamefont {W.}~\bibnamefont {Nazarewicz}}, \ and\ \bibinfo {author}
  {\bibfnamefont {J.~A.}\ \bibnamefont {Sheikh}},\ }\bibfield  {title}
  {\enquote {\bibinfo {title} {Spontaneous fission lifetimes from the
  minimization of self-consistent collective action},}\ }\href {\doibase
  10.1103/PhysRevC.88.064314} {\bibfield  {journal} {\bibinfo  {journal} {Phys.
  Rev. C}\ }\textbf {\bibinfo {volume} {88}},\ \bibinfo {pages} {064314}
  (\bibinfo {year} {2013})}\BibitemShut {NoStop}%
\bibitem [{\citenamefont {Scamps}\ and\ \citenamefont
  {Simenel}(2018)}]{scamps2018a}%
  \BibitemOpen
  \bibfield  {author} {\bibinfo {author} {\bibfnamefont {G.}~\bibnamefont
  {Scamps}}\ and\ \bibinfo {author} {\bibfnamefont {C.}~\bibnamefont
  {Simenel}},\ }\bibfield  {title} {\enquote {\bibinfo {title} {Impact of
  pear-shaped fission fragments on mass-asymmetric fission in actinides},}\
  }\href {\doibase 10.1038/s41586-018-0780-0} {\bibfield  {journal} {\bibinfo
  {journal} {Nature}\ }\textbf {\bibinfo {volume} {564}},\ \bibinfo {pages}
  {382--385} (\bibinfo {year} {2018})}\BibitemShut {NoStop}%
\bibitem [{\citenamefont {Strutinsky}\ and\ \citenamefont
  {Magner}(1976)}]{Strutinsky1976}%
  \BibitemOpen
  \bibfield  {author} {\bibinfo {author} {\bibfnamefont {V.}~\bibnamefont
  {Strutinsky}}\ and\ \bibinfo {author} {\bibfnamefont {A.}~\bibnamefont
  {Magner}},\ }\bibfield  {title} {\enquote {\bibinfo {title} {The
  semiclassical theory of the shell structure of the nucleus},}\ }\href@noop {}
  {\bibfield  {journal} {\bibinfo  {journal} {Sov. J. Part. Nucl.}\ }\textbf
  {\bibinfo {volume} {7}},\ \bibinfo {pages} {459} (\bibinfo {year}
  {1976})}\BibitemShut {NoStop}%
\bibitem [{\citenamefont {ichiro Arita}\ \emph {et~al.}(2020)\citenamefont
  {ichiro Arita}, \citenamefont {Ichikawa},\ and\ \citenamefont
  {Matsuyanagi}}]{Arita2020}%
  \BibitemOpen
  \bibfield  {author} {\bibinfo {author} {\bibfnamefont {K.}~\bibnamefont
  {ichiro Arita}}, \bibinfo {author} {\bibfnamefont {T.}~\bibnamefont
  {Ichikawa}}, \ and\ \bibinfo {author} {\bibfnamefont {K.}~\bibnamefont
  {Matsuyanagi}},\ }\bibfield  {title} {\enquote {\bibinfo {title}
  {Semiclassical origin of asymmetric nuclear fission: nascent-fragment shell
  effect in periodic-orbit theory},}\ }\href {\doibase
  10.1088/1402-4896/ab42a8} {\bibfield  {journal} {\bibinfo  {journal} {Phys.
  Scr.}\ }\textbf {\bibinfo {volume} {95}},\ \bibinfo {pages} {024003}
  (\bibinfo {year} {2020})}\BibitemShut {NoStop}%
\bibitem [{\citenamefont {Bondorf}\ \emph {et~al.}(1995)\citenamefont
  {Bondorf}, \citenamefont {Botvina}, \citenamefont {Iljinov}, \citenamefont
  {Mishustin},\ and\ \citenamefont {Sneppen}}]{Bondorf1995}%
  \BibitemOpen
  \bibfield  {author} {\bibinfo {author} {\bibfnamefont {J.}~\bibnamefont
  {Bondorf}}, \bibinfo {author} {\bibfnamefont {A.}~\bibnamefont {Botvina}},
  \bibinfo {author} {\bibfnamefont {A.}~\bibnamefont {Iljinov}}, \bibinfo
  {author} {\bibfnamefont {I.}~\bibnamefont {Mishustin}}, \ and\ \bibinfo
  {author} {\bibfnamefont {K.}~\bibnamefont {Sneppen}},\ }\bibfield  {title}
  {\enquote {\bibinfo {title} {Statistical multifragmentation of nuclei},}\
  }\href {\doibase https://doi.org/10.1016/0370-1573(94)00097-M} {\bibfield
  {journal} {\bibinfo  {journal} {Phys. Rep.}\ }\textbf {\bibinfo {volume}
  {257}},\ \bibinfo {pages} {133 -- 221} (\bibinfo {year} {1995})}\BibitemShut
  {NoStop}%
\bibitem [{\citenamefont {Simenel}\ and\ \citenamefont {Umar}(2014)}]{Sim14}%
  \BibitemOpen
  \bibfield  {author} {\bibinfo {author} {\bibfnamefont {C.}~\bibnamefont
  {Simenel}}\ and\ \bibinfo {author} {\bibfnamefont {A.~S.}\ \bibnamefont
  {Umar}},\ }\bibfield  {title} {\enquote {\bibinfo {title} {Formation and
  dynamics of fission fragments},}\ }\href {\doibase
  10.1103/PhysRevC.89.031601} {\bibfield  {journal} {\bibinfo  {journal} {Phys.
  Rev. C}\ }\textbf {\bibinfo {volume} {89}},\ \bibinfo {pages} {031601}
  (\bibinfo {year} {2014})}\BibitemShut {NoStop}%
\bibitem [{\citenamefont {Mosel}\ and\ \citenamefont {Schmitt}(1971)}]{Mos71}%
  \BibitemOpen
  \bibfield  {author} {\bibinfo {author} {\bibfnamefont {U.}~\bibnamefont
  {Mosel}}\ and\ \bibinfo {author} {\bibfnamefont {H.}~\bibnamefont
  {Schmitt}},\ }\bibfield  {title} {\enquote {\bibinfo {title} {Potential
  energy surfaces for heavy nuclei in the two-center model},}\ }\href {\doibase
  https://doi.org/10.1016/0375-9474(71)90150-3} {\bibfield  {journal} {\bibinfo
   {journal} {Nucl. Phys. A}\ }\textbf {\bibinfo {volume} {165}},\ \bibinfo
  {pages} {73 -- 96} (\bibinfo {year} {1971})}\BibitemShut {NoStop}%
\bibitem [{\citenamefont {Staszczak}\ \emph {et~al.}(2009)\citenamefont
  {Staszczak}, \citenamefont {Baran}, \citenamefont {Dobaczewski},\ and\
  \citenamefont {Nazarewicz}}]{staszczak2009}%
  \BibitemOpen
  \bibfield  {author} {\bibinfo {author} {\bibfnamefont {A.}~\bibnamefont
  {Staszczak}}, \bibinfo {author} {\bibfnamefont {A.}~\bibnamefont {Baran}},
  \bibinfo {author} {\bibfnamefont {J.}~\bibnamefont {Dobaczewski}}, \ and\
  \bibinfo {author} {\bibfnamefont {W.}~\bibnamefont {Nazarewicz}},\ }\bibfield
   {title} {\enquote {\bibinfo {title} {Microscopic description of complex
  nuclear decay: Multimodal fission},}\ }\href {\doibase
  10.1103/PhysRevC.80.014309} {\bibfield  {journal} {\bibinfo  {journal} {Phys.
  Rev. C}\ }\textbf {\bibinfo {volume} {80}},\ \bibinfo {pages} {014309}
  (\bibinfo {year} {2009})}\BibitemShut {NoStop}%
\bibitem [{\citenamefont {Schunck}\ \emph {et~al.}(2017)\citenamefont
  {Schunck}, \citenamefont {Dobaczewski}, \citenamefont {Satu{\l}a},
  \citenamefont {B{\c{a}}czyk}, \citenamefont {Dudek}, \citenamefont {Gao},
  \citenamefont {Konieczka}, \citenamefont {Sato}, \citenamefont {Shi},
  \citenamefont {Wang},\ and\ \citenamefont {Werner}}]{schunck2017}%
  \BibitemOpen
  \bibfield  {author} {\bibinfo {author} {\bibfnamefont {N.}~\bibnamefont
  {Schunck}}, \bibinfo {author} {\bibfnamefont {J.}~\bibnamefont
  {Dobaczewski}}, \bibinfo {author} {\bibfnamefont {W.}~\bibnamefont
  {Satu{\l}a}}, \bibinfo {author} {\bibfnamefont {P.}~\bibnamefont
  {B{\c{a}}czyk}}, \bibinfo {author} {\bibfnamefont {J.}~\bibnamefont {Dudek}},
  \bibinfo {author} {\bibfnamefont {Y.}~\bibnamefont {Gao}}, \bibinfo {author}
  {\bibfnamefont {M.}~\bibnamefont {Konieczka}}, \bibinfo {author}
  {\bibfnamefont {K.}~\bibnamefont {Sato}}, \bibinfo {author} {\bibfnamefont
  {Y.}~\bibnamefont {Shi}}, \bibinfo {author} {\bibfnamefont {X.}~\bibnamefont
  {Wang}}, \ and\ \bibinfo {author} {\bibfnamefont {T.}~\bibnamefont
  {Werner}},\ }\bibfield  {title} {\enquote {\bibinfo {title} {{Solution of the
  Skyrme-Hartree-Fock-Bogolyubov equations in the Cartesian deformed
  harmonic-oscillator basis. (VIII) HFODD (v2.73y): A new version of the
  program}},}\ }\href {\doibase 10.1016/j.cpc.2017.03.007} {\bibfield
  {journal} {\bibinfo  {journal} {Comput. Phys. Commun.}\ }\textbf {\bibinfo
  {volume} {216}},\ \bibinfo {pages} {145--174} (\bibinfo {year}
  {2017})}\BibitemShut {NoStop}%
\bibitem [{\citenamefont {Bartel}\ \emph {et~al.}(1982)\citenamefont {Bartel},
  \citenamefont {Quentin}, \citenamefont {Brack}, \citenamefont {Guet},\ and\
  \citenamefont {H{\aa}kansson}}]{bartel1982}%
  \BibitemOpen
  \bibfield  {author} {\bibinfo {author} {\bibfnamefont {J.}~\bibnamefont
  {Bartel}}, \bibinfo {author} {\bibfnamefont {P.}~\bibnamefont {Quentin}},
  \bibinfo {author} {\bibfnamefont {M.}~\bibnamefont {Brack}}, \bibinfo
  {author} {\bibfnamefont {C.}~\bibnamefont {Guet}}, \ and\ \bibinfo {author}
  {\bibfnamefont {H.-B.}\ \bibnamefont {H{\aa}kansson}},\ }\bibfield  {title}
  {\enquote {\bibinfo {title} {{Towards a better parametrisation of Skyrme-like
  effective forces: A critical study of the SkM force}},}\ }\href {\doibase
  10.1016/0375-9474(82)90403-1} {\bibfield  {journal} {\bibinfo  {journal}
  {Nucl. Phys. A}\ }\textbf {\bibinfo {volume} {386}},\ \bibinfo {pages}
  {79--100} (\bibinfo {year} {1982})}\BibitemShut {NoStop}%
\bibitem [{\citenamefont {Schunck}\ \emph
  {et~al.}(2015{\natexlab{a}})\citenamefont {Schunck}, \citenamefont
  {McDonnell}, \citenamefont {Sarich}, \citenamefont {Wild},\ and\
  \citenamefont {Higdon}}]{schunck2015}%
  \BibitemOpen
  \bibfield  {author} {\bibinfo {author} {\bibfnamefont {N.}~\bibnamefont
  {Schunck}}, \bibinfo {author} {\bibfnamefont {J.~D.}\ \bibnamefont
  {McDonnell}}, \bibinfo {author} {\bibfnamefont {J.}~\bibnamefont {Sarich}},
  \bibinfo {author} {\bibfnamefont {S.~M.}\ \bibnamefont {Wild}}, \ and\
  \bibinfo {author} {\bibfnamefont {D.}~\bibnamefont {Higdon}},\ }\bibfield
  {title} {\enquote {\bibinfo {title} {Error analysis in nuclear density
  functional theory},}\ }\href {\doibase 10.1088/0954-3899/42/3/034024}
  {\bibfield  {journal} {\bibinfo  {journal} {J. Phys. G}\ }\textbf {\bibinfo
  {volume} {42}},\ \bibinfo {pages} {034024} (\bibinfo {year}
  {2015}{\natexlab{a}})}\BibitemShut {NoStop}%
\bibitem [{\citenamefont {Dobaczewski}\ \emph {et~al.}(2002)\citenamefont
  {Dobaczewski}, \citenamefont {Nazarewicz},\ and\ \citenamefont
  {Stoitsov}}]{dobaczewski2002}%
  \BibitemOpen
  \bibfield  {author} {\bibinfo {author} {\bibfnamefont {J.}~\bibnamefont
  {Dobaczewski}}, \bibinfo {author} {\bibfnamefont {W.}~\bibnamefont
  {Nazarewicz}}, \ and\ \bibinfo {author} {\bibfnamefont {M.}~\bibnamefont
  {Stoitsov}},\ }\bibfield  {title} {\enquote {\bibinfo {title} {Nuclear
  ground-state properties from mean-field calculations},}\ }\href {\doibase
  10.1140/epja/i2001-10218-8} {\bibfield  {journal} {\bibinfo  {journal} {Eur.
  Phys. J. A}\ }\textbf {\bibinfo {volume} {15}},\ \bibinfo {pages} {21--26}
  (\bibinfo {year} {2002})}\BibitemShut {NoStop}%
\bibitem [{\citenamefont {Pei}\ \emph {et~al.}(2009)\citenamefont {Pei},
  \citenamefont {Nazarewicz}, \citenamefont {Sheikh},\ and\ \citenamefont
  {Kerman}}]{Pei09}%
  \BibitemOpen
  \bibfield  {author} {\bibinfo {author} {\bibfnamefont {J.~C.}\ \bibnamefont
  {Pei}}, \bibinfo {author} {\bibfnamefont {W.}~\bibnamefont {Nazarewicz}},
  \bibinfo {author} {\bibfnamefont {J.~A.}\ \bibnamefont {Sheikh}}, \ and\
  \bibinfo {author} {\bibfnamefont {A.~K.}\ \bibnamefont {Kerman}},\ }\bibfield
   {title} {\enquote {\bibinfo {title} {Fission barriers of compound superheavy
  nuclei},}\ }\href {\doibase 10.1103/PhysRevLett.102.192501} {\bibfield
  {journal} {\bibinfo  {journal} {Phys. Rev. Lett.}\ }\textbf {\bibinfo
  {volume} {102}},\ \bibinfo {pages} {192501} (\bibinfo {year}
  {2009})}\BibitemShut {NoStop}%
\bibitem [{\citenamefont {Martin}\ and\ \citenamefont
  {Robledo}(2009)}]{martin2009}%
  \BibitemOpen
  \bibfield  {author} {\bibinfo {author} {\bibfnamefont {V.}~\bibnamefont
  {Martin}}\ and\ \bibinfo {author} {\bibfnamefont {L.~M.}\ \bibnamefont
  {Robledo}},\ }\bibfield  {title} {\enquote {\bibinfo {title} {Fission
  barriers at finite temperature: A theoretical description with the {Gogny}
  force},}\ }\href {\doibase 10.1142/S0218301309012963} {\bibfield  {journal}
  {\bibinfo  {journal} {Int. J. Mod. Phys. E}\ }\textbf {\bibinfo {volume}
  {18}},\ \bibinfo {pages} {861--868} (\bibinfo {year} {2009})}\BibitemShut
  {NoStop}%
\bibitem [{\citenamefont {Schunck}\ \emph
  {et~al.}(2015{\natexlab{b}})\citenamefont {Schunck}, \citenamefont {Duke},\
  and\ \citenamefont {Carr}}]{Sch15}%
  \BibitemOpen
  \bibfield  {author} {\bibinfo {author} {\bibfnamefont {N.}~\bibnamefont
  {Schunck}}, \bibinfo {author} {\bibfnamefont {D.}~\bibnamefont {Duke}}, \
  and\ \bibinfo {author} {\bibfnamefont {H.}~\bibnamefont {Carr}},\ }\bibfield
  {title} {\enquote {\bibinfo {title} {Description of induced nuclear fission
  with skyrme energy functionals. ii. finite temperature effects},}\ }\href
  {\doibase 10.1103/PhysRevC.91.034327} {\bibfield  {journal} {\bibinfo
  {journal} {Phys. Rev. C}\ }\textbf {\bibinfo {volume} {91}},\ \bibinfo
  {pages} {034327} (\bibinfo {year} {2015}{\natexlab{b}})}\BibitemShut
  {NoStop}%
\bibitem [{\citenamefont {Reinhard}\ \emph {et~al.}(2011)\citenamefont
  {Reinhard}, \citenamefont {Maruhn}, \citenamefont {Umar},\ and\ \citenamefont
  {Oberacker}}]{reinhard2011}%
  \BibitemOpen
  \bibfield  {author} {\bibinfo {author} {\bibfnamefont {P.-G.}\ \bibnamefont
  {Reinhard}}, \bibinfo {author} {\bibfnamefont {J.~A.}\ \bibnamefont
  {Maruhn}}, \bibinfo {author} {\bibfnamefont {A.~S.}\ \bibnamefont {Umar}}, \
  and\ \bibinfo {author} {\bibfnamefont {V.~E.}\ \bibnamefont {Oberacker}},\
  }\bibfield  {title} {\enquote {\bibinfo {title} {Localization in light
  nuclei},}\ }\href {\doibase 10.1103/PhysRevC.83.034312} {\bibfield  {journal}
  {\bibinfo  {journal} {Phys. Rev. C}\ }\textbf {\bibinfo {volume} {83}},\
  \bibinfo {pages} {034312} (\bibinfo {year} {2011})}\BibitemShut {NoStop}%
\bibitem [{\citenamefont {Zhang}\ \emph {et~al.}(2016)\citenamefont {Zhang},
  \citenamefont {Schuetrumpf},\ and\ \citenamefont {Nazarewicz}}]{zhang2016}%
  \BibitemOpen
  \bibfield  {author} {\bibinfo {author} {\bibfnamefont {C.~L.}\ \bibnamefont
  {Zhang}}, \bibinfo {author} {\bibfnamefont {B.}~\bibnamefont {Schuetrumpf}},
  \ and\ \bibinfo {author} {\bibfnamefont {W.}~\bibnamefont {Nazarewicz}},\
  }\bibfield  {title} {\enquote {\bibinfo {title} {Nucleon localization and
  fragment formation in nuclear fission},}\ }\href {\doibase
  10.1103/PhysRevC.94.064323} {\bibfield  {journal} {\bibinfo  {journal} {Phys.
  Rev. C}\ }\textbf {\bibinfo {volume} {94}},\ \bibinfo {pages} {064323}
  (\bibinfo {year} {2016})}\BibitemShut {NoStop}%
\bibitem [{\citenamefont {Myers}\ and\ \citenamefont
  {Swiatecki}(1966)}]{Mye66}%
  \BibitemOpen
  \bibfield  {author} {\bibinfo {author} {\bibfnamefont {W.~D.}\ \bibnamefont
  {Myers}}\ and\ \bibinfo {author} {\bibfnamefont {W.~J.}\ \bibnamefont
  {Swiatecki}},\ }\bibfield  {title} {\enquote {\bibinfo {title} {Nuclear
  masses and deformations},}\ }\href {\doibase
  https://doi.org/10.1016/0029-5582(66)90639-0} {\bibfield  {journal} {\bibinfo
   {journal} {Nucl. Phys.}\ }\textbf {\bibinfo {volume} {81}},\ \bibinfo
  {pages} {1 -- 60} (\bibinfo {year} {1966})}\BibitemShut {NoStop}%
\bibitem [{\citenamefont {Warda}\ \emph {et~al.}(2012)\citenamefont {Warda},
  \citenamefont {Staszczak},\ and\ \citenamefont {Nazarewicz}}]{warda2012a}%
  \BibitemOpen
  \bibfield  {author} {\bibinfo {author} {\bibfnamefont {M.}~\bibnamefont
  {Warda}}, \bibinfo {author} {\bibfnamefont {A.}~\bibnamefont {Staszczak}}, \
  and\ \bibinfo {author} {\bibfnamefont {W.}~\bibnamefont {Nazarewicz}},\
  }\bibfield  {title} {\enquote {\bibinfo {title} {Fission modes of mercury
  isotopes},}\ }\href {\doibase 10.1103/PhysRevC.86.024601} {\bibfield
  {journal} {\bibinfo  {journal} {Phys. Rev. C}\ }\textbf {\bibinfo {volume}
  {86}},\ \bibinfo {pages} {024601} (\bibinfo {year} {2012})}\BibitemShut
  {NoStop}%
\bibitem [{\citenamefont {Tsekhanovich}\ \emph {et~al.}(2019)\citenamefont
  {Tsekhanovich}, \citenamefont {Andreyev}, \citenamefont {Nishio},
  \citenamefont {Denis-Petit}, \citenamefont {Hirose}, \citenamefont {Makii},
  \citenamefont {Matheson}, \citenamefont {Morimoto}, \citenamefont {Morita},
  \citenamefont {Nazarewicz}, \citenamefont {Orlandi}, \citenamefont
  {Sadhukhan}, \citenamefont {Tanaka}, \citenamefont {Vermeulen},\ and\
  \citenamefont {Warda}}]{tsekhanovich2019}%
  \BibitemOpen
  \bibfield  {author} {\bibinfo {author} {\bibfnamefont {I.}~\bibnamefont
  {Tsekhanovich}}, \bibinfo {author} {\bibfnamefont {A.}~\bibnamefont
  {Andreyev}}, \bibinfo {author} {\bibfnamefont {K.}~\bibnamefont {Nishio}},
  \bibinfo {author} {\bibfnamefont {D.}~\bibnamefont {Denis-Petit}}, \bibinfo
  {author} {\bibfnamefont {K.}~\bibnamefont {Hirose}}, \bibinfo {author}
  {\bibfnamefont {H.}~\bibnamefont {Makii}}, \bibinfo {author} {\bibfnamefont
  {Z.}~\bibnamefont {Matheson}}, \bibinfo {author} {\bibfnamefont
  {K.}~\bibnamefont {Morimoto}}, \bibinfo {author} {\bibfnamefont
  {K.}~\bibnamefont {Morita}}, \bibinfo {author} {\bibfnamefont
  {W.}~\bibnamefont {Nazarewicz}}, \bibinfo {author} {\bibfnamefont
  {R.}~\bibnamefont {Orlandi}}, \bibinfo {author} {\bibfnamefont
  {J.}~\bibnamefont {Sadhukhan}}, \bibinfo {author} {\bibfnamefont
  {T.}~\bibnamefont {Tanaka}}, \bibinfo {author} {\bibfnamefont
  {M.}~\bibnamefont {Vermeulen}}, \ and\ \bibinfo {author} {\bibfnamefont
  {M.}~\bibnamefont {Warda}},\ }\bibfield  {title} {\enquote {\bibinfo {title}
  {Observation of the competing fission modes in $^{178}${Pt}},}\ }\href
  {\doibase 10.1016/j.physletb.2019.02.006} {\bibfield  {journal} {\bibinfo
  {journal} {Phys. Lett. B}\ }\textbf {\bibinfo {volume} {790}},\ \bibinfo
  {pages} {583--588} (\bibinfo {year} {2019})}\BibitemShut {NoStop}%
\bibitem [{\citenamefont {Hulet}\ \emph {et~al.}(1989)\citenamefont {Hulet},
  \citenamefont {Wild}, \citenamefont {Dougan}, \citenamefont {Lougheed},
  \citenamefont {Landrum}, \citenamefont {Dougan}, \citenamefont {Baisden},
  \citenamefont {Henderson}, \citenamefont {Dupzyk}, \citenamefont {Hahn},
  \citenamefont {Sch{\"{a}}del}, \citenamefont {S{\"{u}}mmerer},\ and\
  \citenamefont {Bethune}}]{hulet1989}%
  \BibitemOpen
  \bibfield  {author} {\bibinfo {author} {\bibfnamefont {E.~K.}\ \bibnamefont
  {Hulet}}, \bibinfo {author} {\bibfnamefont {J.~F.}\ \bibnamefont {Wild}},
  \bibinfo {author} {\bibfnamefont {R.~J.}\ \bibnamefont {Dougan}}, \bibinfo
  {author} {\bibfnamefont {R.~W.}\ \bibnamefont {Lougheed}}, \bibinfo {author}
  {\bibfnamefont {J.~H.}\ \bibnamefont {Landrum}}, \bibinfo {author}
  {\bibfnamefont {A.~D.}\ \bibnamefont {Dougan}}, \bibinfo {author}
  {\bibfnamefont {P.~A.}\ \bibnamefont {Baisden}}, \bibinfo {author}
  {\bibfnamefont {C.~M.}\ \bibnamefont {Henderson}}, \bibinfo {author}
  {\bibfnamefont {R.~J.}\ \bibnamefont {Dupzyk}}, \bibinfo {author}
  {\bibfnamefont {R.~L.}\ \bibnamefont {Hahn}}, \bibinfo {author}
  {\bibfnamefont {M.}~\bibnamefont {Sch{\"{a}}del}}, \bibinfo {author}
  {\bibfnamefont {K.}~\bibnamefont {S{\"{u}}mmerer}}, \ and\ \bibinfo {author}
  {\bibfnamefont {G.~R.}\ \bibnamefont {Bethune}},\ }\bibfield  {title}
  {\enquote {\bibinfo {title} {Spontaneous fission properties of
  $^{258}\mathrm{Fm}$, $^{259}\mathrm{Md}$, $^{260}\mathrm{Md}$,
  $^{258}\mathrm{No}$, and $^{260}]$: Bimodal fission},}\ }\href {\doibase
  10.1103/PhysRevC.40.770} {\bibfield  {journal} {\bibinfo  {journal} {Phys.
  Rev. C}\ }\textbf {\bibinfo {volume} {40}},\ \bibinfo {pages} {770--784}
  (\bibinfo {year} {1989})}\BibitemShut {NoStop}%
\bibitem [{\citenamefont {M{\"o}ller}\ \emph {et~al.}(1989)\citenamefont
  {M{\"o}ller}, \citenamefont {Nix},\ and\ \citenamefont
  {Swiatecki}}]{moller1989}%
  \BibitemOpen
  \bibfield  {author} {\bibinfo {author} {\bibfnamefont {P.}~\bibnamefont
  {M{\"o}ller}}, \bibinfo {author} {\bibfnamefont {J.}~\bibnamefont {Nix}}, \
  and\ \bibinfo {author} {\bibfnamefont {W.}~\bibnamefont {Swiatecki}},\
  }\bibfield  {title} {\enquote {\bibinfo {title} {New developments in the
  calculation of heavy-element fission barriers},}\ }\href {\doibase
  10.1016/0375-9474(89)90403-X} {\bibfield  {journal} {\bibinfo  {journal}
  {Nucl. Phys. A}\ }\textbf {\bibinfo {volume} {492}},\ \bibinfo {pages} {349
  -- 387} (\bibinfo {year} {1989})}\BibitemShut {NoStop}%
\bibitem [{\citenamefont {Brosa}\ \emph {et~al.}(1990)\citenamefont {Brosa},
  \citenamefont {Grossmann},\ and\ \citenamefont {M{\"u}ller}}]{Brosa90}%
  \BibitemOpen
  \bibfield  {author} {\bibinfo {author} {\bibfnamefont {U.}~\bibnamefont
  {Brosa}}, \bibinfo {author} {\bibfnamefont {S.}~\bibnamefont {Grossmann}}, \
  and\ \bibinfo {author} {\bibfnamefont {A.}~\bibnamefont {M{\"u}ller}},\
  }\bibfield  {title} {\enquote {\bibinfo {title} {Nuclear scission},}\ }\href
  {\doibase 10.1016/0370-1573(90)90114-H} {\bibfield  {journal} {\bibinfo
  {journal} {Phys. Rep.}\ }\textbf {\bibinfo {volume} {197}},\ \bibinfo {pages}
  {167 -- 262} (\bibinfo {year} {1990})}\BibitemShut {NoStop}%
\bibitem [{\citenamefont {Sadhukhan}\ \emph {et~al.}(2014)\citenamefont
  {Sadhukhan}, \citenamefont {Dobaczewski}, \citenamefont {Nazarewicz},
  \citenamefont {Sheikh},\ and\ \citenamefont {Baran}}]{Sadhukhan2014}%
  \BibitemOpen
  \bibfield  {author} {\bibinfo {author} {\bibfnamefont {J.}~\bibnamefont
  {Sadhukhan}}, \bibinfo {author} {\bibfnamefont {J.}~\bibnamefont
  {Dobaczewski}}, \bibinfo {author} {\bibfnamefont {W.}~\bibnamefont
  {Nazarewicz}}, \bibinfo {author} {\bibfnamefont {J.~A.}\ \bibnamefont
  {Sheikh}}, \ and\ \bibinfo {author} {\bibfnamefont {A.}~\bibnamefont
  {Baran}},\ }\bibfield  {title} {\enquote {\bibinfo {title} {Pairing-induced
  speedup of nuclear spontaneous fission},}\ }\href {\doibase
  10.1103/PhysRevC.90.061304} {\bibfield  {journal} {\bibinfo  {journal} {Phys.
  Rev. C}\ }\textbf {\bibinfo {volume} {90}},\ \bibinfo {pages} {061304}
  (\bibinfo {year} {2014})}\BibitemShut {NoStop}%
\bibitem [{\citenamefont {Zhao}\ \emph {et~al.}(2015)\citenamefont {Zhao},
  \citenamefont {Lu}, \citenamefont {Nik{\v{s}}i{\'{c}}},\ and\ \citenamefont
  {Vretenar}}]{zhao2015}%
  \BibitemOpen
  \bibfield  {author} {\bibinfo {author} {\bibfnamefont {J.}~\bibnamefont
  {Zhao}}, \bibinfo {author} {\bibfnamefont {B.-N.}\ \bibnamefont {Lu}},
  \bibinfo {author} {\bibfnamefont {T.}~\bibnamefont {Nik{\v{s}}i{\'{c}}}}, \
  and\ \bibinfo {author} {\bibfnamefont {D.}~\bibnamefont {Vretenar}},\
  }\bibfield  {title} {\enquote {\bibinfo {title} {Multidimensionally
  constrained relativistic {Hartree-Bogoliubov} study of spontaneous nuclear
  fission},}\ }\href {\doibase 10.1103/PhysRevC.92.064315} {\bibfield
  {journal} {\bibinfo  {journal} {Phys. Rev. C}\ }\textbf {\bibinfo {volume}
  {92}},\ \bibinfo {pages} {064315} (\bibinfo {year} {2015})}\BibitemShut
  {NoStop}%
\bibitem [{\citenamefont {Baran}\ \emph {et~al.}(2011)\citenamefont {Baran},
  \citenamefont {Sheikh}, \citenamefont {Dobaczewski}, \citenamefont
  {Nazarewicz},\ and\ \citenamefont {Staszczak}}]{Baran2011}%
  \BibitemOpen
  \bibfield  {author} {\bibinfo {author} {\bibfnamefont {A.}~\bibnamefont
  {Baran}}, \bibinfo {author} {\bibfnamefont {J.~A.}\ \bibnamefont {Sheikh}},
  \bibinfo {author} {\bibfnamefont {J.}~\bibnamefont {Dobaczewski}}, \bibinfo
  {author} {\bibfnamefont {W.}~\bibnamefont {Nazarewicz}}, \ and\ \bibinfo
  {author} {\bibfnamefont {A.}~\bibnamefont {Staszczak}},\ }\bibfield  {title}
  {\enquote {\bibinfo {title} {Quadrupole collective inertia in nuclear
  fission: Cranking approximation},}\ }\href {\doibase
  10.1103/PhysRevC.84.054321} {\bibfield  {journal} {\bibinfo  {journal} {Phys.
  Rev. C}\ }\textbf {\bibinfo {volume} {84}},\ \bibinfo {pages} {054321}
  (\bibinfo {year} {2011})}\BibitemShut {NoStop}%
\bibitem [{\citenamefont {Giuliani}\ and\ \citenamefont
  {Robledo}(2018)}]{giuliani2018b}%
  \BibitemOpen
  \bibfield  {author} {\bibinfo {author} {\bibfnamefont {S.~A.}\ \bibnamefont
  {Giuliani}}\ and\ \bibinfo {author} {\bibfnamefont {L.~M.}\ \bibnamefont
  {Robledo}},\ }\bibfield  {title} {\enquote {\bibinfo {title}
  {Non-perturbative collective inertias for fission: A comparative study},}\
  }\href {\doibase 10.1016/j.physletb.2018.10.045} {\bibfield  {journal}
  {\bibinfo  {journal} {Phys. Lett. B}\ }\textbf {\bibinfo {volume} {787}},\
  \bibinfo {pages} {134--140} (\bibinfo {year} {2018})}\BibitemShut {NoStop}%
\bibitem [{\citenamefont {Moretto}\ and\ \citenamefont
  {Babinet}(1974)}]{(mor74)}%
  \BibitemOpen
  \bibfield  {author} {\bibinfo {author} {\bibfnamefont {L.}~\bibnamefont
  {Moretto}}\ and\ \bibinfo {author} {\bibfnamefont {R.}~\bibnamefont
  {Babinet}},\ }\bibfield  {title} {\enquote {\bibinfo {title} {Large
  superfluidity enhancement in the penetration of the fission barrier},}\
  }\href {\doibase 10.1016/0370-2693(74)90494-8} {\bibfield  {journal}
  {\bibinfo  {journal} {Phys. Lett. B}\ }\textbf {\bibinfo {volume} {49}},\
  \bibinfo {pages} {147 -- 149} (\bibinfo {year} {1974})}\BibitemShut {NoStop}%
\bibitem [{\citenamefont {Bernard}\ \emph {et~al.}(2019)\citenamefont
  {Bernard}, \citenamefont {Giuliani},\ and\ \citenamefont
  {Robledo}}]{Bernard2019}%
  \BibitemOpen
  \bibfield  {author} {\bibinfo {author} {\bibfnamefont {R.}~\bibnamefont
  {Bernard}}, \bibinfo {author} {\bibfnamefont {S.~A.}\ \bibnamefont
  {Giuliani}}, \ and\ \bibinfo {author} {\bibfnamefont {L.~M.}\ \bibnamefont
  {Robledo}},\ }\bibfield  {title} {\enquote {\bibinfo {title} {Role of dynamic
  pairing correlations in fission dynamics},}\ }\href {\doibase
  10.1103/PhysRevC.99.064301} {\bibfield  {journal} {\bibinfo  {journal} {Phys.
  Rev. C}\ }\textbf {\bibinfo {volume} {99}},\ \bibinfo {pages} {064301}
  (\bibinfo {year} {2019})}\BibitemShut {NoStop}%
\bibitem [{\citenamefont {Scamps}\ and\ \citenamefont
  {Simenel}(2019)}]{scamps2019a}%
  \BibitemOpen
  \bibfield  {author} {\bibinfo {author} {\bibfnamefont {G.}~\bibnamefont
  {Scamps}}\ and\ \bibinfo {author} {\bibfnamefont {C.}~\bibnamefont
  {Simenel}},\ }\bibfield  {title} {\enquote {\bibinfo {title} {Effect of shell
  structure on the fission of sub-lead nuclei},}\ }\href {\doibase
  10.1103/PhysRevC.100.041602} {\bibfield  {journal} {\bibinfo  {journal}
  {Phys. Rev. C}\ }\textbf {\bibinfo {volume} {100}},\ \bibinfo {pages}
  {041602} (\bibinfo {year} {2019})}\BibitemShut {NoStop}%
\bibitem [{\citenamefont {Laidler}\ and\ \citenamefont
  {Brown}(1962)}]{Laidler1962}%
  \BibitemOpen
  \bibfield  {author} {\bibinfo {author} {\bibfnamefont {J.}~\bibnamefont
  {Laidler}}\ and\ \bibinfo {author} {\bibfnamefont {F.}~\bibnamefont
  {Brown}},\ }\bibfield  {title} {\enquote {\bibinfo {title} {Mass distribution
  in the spontaneous fission of $^{240}${Pu}},}\ }\href {\doibase
  10.1016/0022-1902(62)80001-3} {\bibfield  {journal} {\bibinfo  {journal} {J.
  Inorg. Nucl. Chem.}\ }\textbf {\bibinfo {volume} {24}},\ \bibinfo {pages}
  {1485--1492} (\bibinfo {year} {1962})}\BibitemShut {NoStop}%
\bibitem [{\citenamefont {Brandt}\ \emph {et~al.}(1963)\citenamefont {Brandt},
  \citenamefont {Thompson}, \citenamefont {Gatti},\ and\ \citenamefont
  {Phillips}}]{brandt1963}%
  \BibitemOpen
  \bibfield  {author} {\bibinfo {author} {\bibfnamefont {R.}~\bibnamefont
  {Brandt}}, \bibinfo {author} {\bibfnamefont {S.~G.}\ \bibnamefont
  {Thompson}}, \bibinfo {author} {\bibfnamefont {R.~C.}\ \bibnamefont {Gatti}},
  \ and\ \bibinfo {author} {\bibfnamefont {L.}~\bibnamefont {Phillips}},\
  }\bibfield  {title} {\enquote {\bibinfo {title} {Mass and energy
  distributions in the spontaneous fission of some heavy isotopes},}\ }\href
  {\doibase 10.1103/PhysRev.131.2617} {\bibfield  {journal} {\bibinfo
  {journal} {Phys. Rev.}\ }\textbf {\bibinfo {volume} {131}},\ \bibinfo {pages}
  {2617--2624} (\bibinfo {year} {1963})}\BibitemShut {NoStop}%
\bibitem [{\citenamefont {Gindler}\ \emph {et~al.}(1977)\citenamefont
  {Gindler}, \citenamefont {Flynn}, \citenamefont {Glendenin},\ and\
  \citenamefont {Sjoblom}}]{gindler1977}%
  \BibitemOpen
  \bibfield  {author} {\bibinfo {author} {\bibfnamefont {J.~E.}\ \bibnamefont
  {Gindler}}, \bibinfo {author} {\bibfnamefont {K.~F.}\ \bibnamefont {Flynn}},
  \bibinfo {author} {\bibfnamefont {L.~E.}\ \bibnamefont {Glendenin}}, \ and\
  \bibinfo {author} {\bibfnamefont {R.~K.}\ \bibnamefont {Sjoblom}},\
  }\bibfield  {title} {\enquote {\bibinfo {title} {Distribution of mass,
  kinetic energy, and neutron yield in the spontaneous fission of
  $^{257}${Fm}},}\ }\href {\doibase 10.1103/PhysRevC.16.1483} {\bibfield
  {journal} {\bibinfo  {journal} {Phys. Rev. C}\ }\textbf {\bibinfo {volume}
  {16}},\ \bibinfo {pages} {1483--1492} (\bibinfo {year} {1977})}\BibitemShut
  {NoStop}%
\bibitem [{\citenamefont {Harbour}\ \emph {et~al.}(1973)\citenamefont
  {Harbour}, \citenamefont {MacMurdo}, \citenamefont {Troutner},\ and\
  \citenamefont {Hoehn}}]{harbour1973}%
  \BibitemOpen
  \bibfield  {author} {\bibinfo {author} {\bibfnamefont {R.~M.}\ \bibnamefont
  {Harbour}}, \bibinfo {author} {\bibfnamefont {K.~W.}\ \bibnamefont
  {MacMurdo}}, \bibinfo {author} {\bibfnamefont {D.~E.}\ \bibnamefont
  {Troutner}}, \ and\ \bibinfo {author} {\bibfnamefont {M.~V.}\ \bibnamefont
  {Hoehn}},\ }\bibfield  {title} {\enquote {\bibinfo {title} {Mass and nuclear
  charge distributions from the spontaneous fission of $^{254}${Fm}},}\ }\href
  {\doibase 10.1103/PhysRevC.8.1488} {\bibfield  {journal} {\bibinfo  {journal}
  {Phys. Rev. C}\ }\textbf {\bibinfo {volume} {8}},\ \bibinfo {pages}
  {1488--1493} (\bibinfo {year} {1973})}\BibitemShut {NoStop}%
\bibitem [{\citenamefont {Flynn}\ \emph {et~al.}(1972)\citenamefont {Flynn},
  \citenamefont {Horwitz}, \citenamefont {Bloomquist}, \citenamefont {Barnes},
  \citenamefont {Sjoblom}, \citenamefont {Fields},\ and\ \citenamefont
  {Glendenin}}]{flynn1972}%
  \BibitemOpen
  \bibfield  {author} {\bibinfo {author} {\bibfnamefont {K.~F.}\ \bibnamefont
  {Flynn}}, \bibinfo {author} {\bibfnamefont {E.~P.}\ \bibnamefont {Horwitz}},
  \bibinfo {author} {\bibfnamefont {C.~A.~A.}\ \bibnamefont {Bloomquist}},
  \bibinfo {author} {\bibfnamefont {R.~F.}\ \bibnamefont {Barnes}}, \bibinfo
  {author} {\bibfnamefont {R.~K.}\ \bibnamefont {Sjoblom}}, \bibinfo {author}
  {\bibfnamefont {P.~R.}\ \bibnamefont {Fields}}, \ and\ \bibinfo {author}
  {\bibfnamefont {L.~E.}\ \bibnamefont {Glendenin}},\ }\bibfield  {title}
  {\enquote {\bibinfo {title} {Distribution of mass in the spontaneous fission
  of $^{256}${Fm}},}\ }\href {\doibase 10.1103/PhysRevC.5.1725} {\bibfield
  {journal} {\bibinfo  {journal} {Phys. Rev. C}\ }\textbf {\bibinfo {volume}
  {5}},\ \bibinfo {pages} {1725--1729} (\bibinfo {year} {1972})}\BibitemShut
  {NoStop}%
\bibitem [{\citenamefont {Hoffman}\ \emph {et~al.}(1980)\citenamefont
  {Hoffman}, \citenamefont {Wilhelmy}, \citenamefont {Weber}, \citenamefont
  {Daniels}, \citenamefont {Hulet}, \citenamefont {Lougheed}, \citenamefont
  {Landrum}, \citenamefont {Wild},\ and\ \citenamefont {Dupzyk}}]{hoffman1980}%
  \BibitemOpen
  \bibfield  {author} {\bibinfo {author} {\bibfnamefont {D.~C.}\ \bibnamefont
  {Hoffman}}, \bibinfo {author} {\bibfnamefont {J.~B.}\ \bibnamefont
  {Wilhelmy}}, \bibinfo {author} {\bibfnamefont {J.}~\bibnamefont {Weber}},
  \bibinfo {author} {\bibfnamefont {W.~R.}\ \bibnamefont {Daniels}}, \bibinfo
  {author} {\bibfnamefont {E.~K.}\ \bibnamefont {Hulet}}, \bibinfo {author}
  {\bibfnamefont {R.~W.}\ \bibnamefont {Lougheed}}, \bibinfo {author}
  {\bibfnamefont {J.~H.}\ \bibnamefont {Landrum}}, \bibinfo {author}
  {\bibfnamefont {J.~F.}\ \bibnamefont {Wild}}, \ and\ \bibinfo {author}
  {\bibfnamefont {R.~J.}\ \bibnamefont {Dupzyk}},\ }\bibfield  {title}
  {\enquote {\bibinfo {title} {12.3-min $^{256}\mathrm{Cf}$ and 43-min
  $^{258}\mathrm{Md}$ and systematics of the spontaneous fission properties of
  heavy nuclides},}\ }\href {\doibase 10.1103/PhysRevC.21.972} {\bibfield
  {journal} {\bibinfo  {journal} {Phys. Rev. C}\ }\textbf {\bibinfo {volume}
  {21}},\ \bibinfo {pages} {972--981} (\bibinfo {year} {1980})}\BibitemShut
  {NoStop}%
\bibitem [{\citenamefont {Schmitt}\ \emph {et~al.}(1984)\citenamefont
  {Schmitt}, \citenamefont {Guessous}, \citenamefont {Bocquet}, \citenamefont
  {Clerc}, \citenamefont {Brissot}, \citenamefont {Engelhardt}, \citenamefont
  {Faust}, \citenamefont {Gönnenwein}, \citenamefont {Mutterer}, \citenamefont
  {Nifenecker}, \citenamefont {Pannicke}, \citenamefont {Ristori},\ and\
  \citenamefont {Theobald}}]{schmitt1984}%
  \BibitemOpen
  \bibfield  {author} {\bibinfo {author} {\bibfnamefont {C.}~\bibnamefont
  {Schmitt}}, \bibinfo {author} {\bibfnamefont {A.}~\bibnamefont {Guessous}},
  \bibinfo {author} {\bibfnamefont {J.}~\bibnamefont {Bocquet}}, \bibinfo
  {author} {\bibfnamefont {H.-G.}\ \bibnamefont {Clerc}}, \bibinfo {author}
  {\bibfnamefont {R.}~\bibnamefont {Brissot}}, \bibinfo {author} {\bibfnamefont
  {D.}~\bibnamefont {Engelhardt}}, \bibinfo {author} {\bibfnamefont
  {H.}~\bibnamefont {Faust}}, \bibinfo {author} {\bibfnamefont
  {F.}~\bibnamefont {Gönnenwein}}, \bibinfo {author} {\bibfnamefont
  {M.}~\bibnamefont {Mutterer}}, \bibinfo {author} {\bibfnamefont
  {H.}~\bibnamefont {Nifenecker}}, \bibinfo {author} {\bibfnamefont
  {J.}~\bibnamefont {Pannicke}}, \bibinfo {author} {\bibfnamefont
  {C.}~\bibnamefont {Ristori}}, \ and\ \bibinfo {author} {\bibfnamefont
  {J.}~\bibnamefont {Theobald}},\ }\bibfield  {title} {\enquote {\bibinfo
  {title} {Fission yields at different fission-product kinetic energies for
  thermal-neutron-induced fission of $^{239}$pu},}\ }\href {\doibase
  https://doi.org/10.1016/0375-9474(84)90191-X} {\bibfield  {journal} {\bibinfo
   {journal} {Nucl. Phys. A}\ }\textbf {\bibinfo {volume} {430}},\ \bibinfo
  {pages} {21 -- 60} (\bibinfo {year} {1984})}\BibitemShut {NoStop}%
\bibitem [{\citenamefont {Flynn}\ \emph {et~al.}(1975)\citenamefont {Flynn},
  \citenamefont {Gindler}, \citenamefont {Sjoblom},\ and\ \citenamefont
  {Glendenin}}]{Fly75}%
  \BibitemOpen
  \bibfield  {author} {\bibinfo {author} {\bibfnamefont {K.~F.}\ \bibnamefont
  {Flynn}}, \bibinfo {author} {\bibfnamefont {J.~E.}\ \bibnamefont {Gindler}},
  \bibinfo {author} {\bibfnamefont {R.~K.}\ \bibnamefont {Sjoblom}}, \ and\
  \bibinfo {author} {\bibfnamefont {L.~E.}\ \bibnamefont {Glendenin}},\
  }\bibfield  {title} {\enquote {\bibinfo {title} {Mass distributions for
  thermal-neutron-induced fission of $^{255}\mathrm{Fm}$ and
  $^{251}\mathrm{Cf}$},}\ }\href {\doibase 10.1103/PhysRevC.11.1676} {\bibfield
   {journal} {\bibinfo  {journal} {Phys. Rev. C}\ }\textbf {\bibinfo {volume}
  {11}},\ \bibinfo {pages} {1676--1680} (\bibinfo {year} {1975})}\BibitemShut
  {NoStop}%
\bibitem [{\citenamefont {Schmidt}\ \emph {et~al.}(2008)\citenamefont
  {Schmidt}, \citenamefont {Keli{\'{c}}},\ and\ \citenamefont
  {Ricciardi}}]{schmidt2008}%
  \BibitemOpen
  \bibfield  {author} {\bibinfo {author} {\bibfnamefont {K.-H.}\ \bibnamefont
  {Schmidt}}, \bibinfo {author} {\bibfnamefont {A.}~\bibnamefont
  {Keli{\'{c}}}}, \ and\ \bibinfo {author} {\bibfnamefont {M.~V.}\ \bibnamefont
  {Ricciardi}},\ }\bibfield  {title} {\enquote {\bibinfo {title} {{Experimental
  evidence for the separability of compound-nucleus and fragment properties in
  fission}},}\ }\href {\doibase 10.1209/0295-5075/83/32001} {\bibfield
  {journal} {\bibinfo  {journal} {Europhys. Lett.}\ }\textbf {\bibinfo {volume}
  {83}},\ \bibinfo {pages} {32001} (\bibinfo {year} {2008})}\BibitemShut
  {NoStop}%
\bibitem [{\citenamefont {Goriely}\ \emph {et~al.}(2009)\citenamefont
  {Goriely}, \citenamefont {Hilaire}, \citenamefont {Koning}, \citenamefont
  {Sin},\ and\ \citenamefont {Capote}}]{goriely2009}%
  \BibitemOpen
  \bibfield  {author} {\bibinfo {author} {\bibfnamefont {S.}~\bibnamefont
  {Goriely}}, \bibinfo {author} {\bibfnamefont {S.}~\bibnamefont {Hilaire}},
  \bibinfo {author} {\bibfnamefont {A.~J.}\ \bibnamefont {Koning}}, \bibinfo
  {author} {\bibfnamefont {M.}~\bibnamefont {Sin}}, \ and\ \bibinfo {author}
  {\bibfnamefont {R.}~\bibnamefont {Capote}},\ }\bibfield  {title} {\enquote
  {\bibinfo {title} {{Towards a prediction of fission cross sections on the
  basis of microscopic nuclear inputs}},}\ }\href {\doibase
  10.1103/PhysRevC.79.024612} {\bibfield  {journal} {\bibinfo  {journal} {Phys.
  Rev. C}\ }\textbf {\bibinfo {volume} {79}},\ \bibinfo {pages} {024612}
  (\bibinfo {year} {2009})}\BibitemShut {NoStop}%
\bibitem [{\citenamefont {Erler}\ \emph {et~al.}(2012)\citenamefont {Erler},
  \citenamefont {Langanke}, \citenamefont {Loens}, \citenamefont
  {Mart{\'{i}}nez-Pinedo},\ and\ \citenamefont {Reinhard}}]{erler2012}%
  \BibitemOpen
  \bibfield  {author} {\bibinfo {author} {\bibfnamefont {J.}~\bibnamefont
  {Erler}}, \bibinfo {author} {\bibfnamefont {K.}~\bibnamefont {Langanke}},
  \bibinfo {author} {\bibfnamefont {H.~P.}\ \bibnamefont {Loens}}, \bibinfo
  {author} {\bibfnamefont {G.}~\bibnamefont {Mart{\'{i}}nez-Pinedo}}, \ and\
  \bibinfo {author} {\bibfnamefont {P.-G.}\ \bibnamefont {Reinhard}},\
  }\bibfield  {title} {\enquote {\bibinfo {title} {Fission properties for
  r-process nuclei},}\ }\href {\doibase 10.1103/PhysRevC.85.025802} {\bibfield
  {journal} {\bibinfo  {journal} {Phys. Rev. C}\ }\textbf {\bibinfo {volume}
  {85}},\ \bibinfo {pages} {25802} (\bibinfo {year} {2012})}\BibitemShut
  {NoStop}%
\bibitem [{\citenamefont {Giuliani}\ \emph {et~al.}(2018)\citenamefont
  {Giuliani}, \citenamefont {Mart{\'{i}}nez-Pinedo},\ and\ \citenamefont
  {Robledo}}]{giuliani2018a}%
  \BibitemOpen
  \bibfield  {author} {\bibinfo {author} {\bibfnamefont {S.~A.}\ \bibnamefont
  {Giuliani}}, \bibinfo {author} {\bibfnamefont {G.}~\bibnamefont
  {Mart{\'{i}}nez-Pinedo}}, \ and\ \bibinfo {author} {\bibfnamefont {L.~M.}\
  \bibnamefont {Robledo}},\ }\bibfield  {title} {\enquote {\bibinfo {title}
  {Fission properties of superheavy nuclei for r-process calculations},}\
  }\href {\doibase 10.1103/PhysRevC.97.034323} {\bibfield  {journal} {\bibinfo
  {journal} {Phys. Rev. C}\ }\textbf {\bibinfo {volume} {97}},\ \bibinfo
  {pages} {034323} (\bibinfo {year} {2018})}\BibitemShut {NoStop}%
\bibitem [{\citenamefont {Giuliani}\ \emph {et~al.}(2019)\citenamefont
  {Giuliani}, \citenamefont {Mart{\'{i}}nez-Pinedo}, \citenamefont {Wu},\ and\
  \citenamefont {Robledo}}]{giuliani2019a}%
  \BibitemOpen
  \bibfield  {author} {\bibinfo {author} {\bibfnamefont {S.~A.}\ \bibnamefont
  {Giuliani}}, \bibinfo {author} {\bibfnamefont {G.}~\bibnamefont
  {Mart{\'{i}}nez-Pinedo}}, \bibinfo {author} {\bibfnamefont {M.~R.}\
  \bibnamefont {Wu}}, \ and\ \bibinfo {author} {\bibfnamefont {L.~M.}\
  \bibnamefont {Robledo}},\ }\bibfield  {title} {\enquote {\bibinfo {title}
  {Fission and the r-process nucleosynthesis of translead nuclei},}\ }\href
  {http://arxiv.org/abs/1904.03733} {\  (\bibinfo {year} {2019})},\ \Eprint
  {http://arxiv.org/abs/1904.03733} {arXiv:1904.03733 [nucl-th]} \BibitemShut
  {NoStop}%
\bibitem [{\citenamefont {Gönnenwein}(2013)}]{Gonnenwein2013}%
  \BibitemOpen
  \bibfield  {author} {\bibinfo {author} {\bibfnamefont {F.}~\bibnamefont
  {Gönnenwein}},\ }\bibfield  {title} {\enquote {\bibinfo {title} {Even-odd
  effects of fragment yields in low energy fission},}\ }\href {\doibase
  10.1016/j.phpro.2013.06.016} {\bibfield  {journal} {\bibinfo  {journal}
  {Phys. Procedia}\ }\textbf {\bibinfo {volume} {47}},\ \bibinfo {pages} {107
  -- 114} (\bibinfo {year} {2013})}\BibitemShut {NoStop}%
\end{thebibliography}%

\end{document}